\documentclass[aps,superscriptaddress,nofootinbib,showpacs]{revtex4}
\usepackage{graphicx} 
\usepackage{color}
\usepackage{multirow}
\begin{document}

\title{Non-standard neutrinos interactions in a 331 model with minimum Higgs 
sector}

\author{M. Medina\footnote{mmedina@ifi.unicamp.br}} 
\affiliation{Instituto de F\'isica Gleb Wataghin - UNICAMP, {13083-859}, Campinas SP, Brazil}

\author{P. C. de Holanda\footnote{holanda@ifi.unicamp.br}}
\affiliation{Instituto de F\'isica Gleb Wataghin - UNICAMP, {13083-859}, Campinas SP, Brazil}

\begin{abstract}

We present a detailed analysis of a class of extensions to the
SM Gauge chiral symmetry $SU(3)_{C}\times SU(3)_{L}\times U(1)_{X}$
(331 model), where the neutrino electroweak interaction 
with matter via charged and neutral current is modified through new 
gauge bosons of the model.
We found the connections between the non-standard contributions
on 331 model with non-standard interactions. Through limits of such 
interactions in cross section experiments we constrained the parameters of 
the model, obtaining that the new energy scale of this theory should obey
$V>1.3$ TeV and the new bosons of the model must have masses greater 
than 610 GeV.

\end{abstract}
\pacs{}
\maketitle

\section{INTRODUCTION}

The confirmation of flavor neutrino oscillation~\cite{int-8,int-7}
by the combination of a variety of data from 
solar~\cite{sol-1,sol-2,sol-3,sol-4,sol-5},
atmospheric~\cite{atm-1,atm-2,atm-3,atm-4,atm-5} and 
reactor~\cite{reac-1,reac-2,reac-3,reac-4,reac-5}
neutrino experiments established the incompleteness
of the Standard Model of electroweak interactions, leaving room for other
non-standard neutrino properties.  

One convenient way to describe neutrino new interactions 
with matter in the electro-weak (EW) broken phase are the so-called
non-standard neutrino interactions (NSI), that is a very widespread
and convenient way of parameterizing the effects of new physics in
neutrino oscillations~\cite{nsi-1,nsi-2,nsi-3,nsi-4,nsi-5,nsi-6}.
NSI with first generation of leptons and quarks for four-fermion
operators are contained in the following Lagrangian density\cite{nsi-1,nsi-2,nsi-5}:
\begin{equation}
\mathcal{L}_{eff}^{NSI}=-2\sqrt{2}G_{F}\sum_{f,P}\varepsilon_{\alpha\beta}^{fP}\left[\bar{f}\gamma^{\mu}Pf\right]\left[\bar{\nu_{\alpha}}\gamma^{\mu}L\nu_{\beta}\right],\label{eq:NSI1}
\end{equation}
where $G_{F}$ is the Fermi constant, $f=u,d,e$ and $P=L,R$ with
$2L=\left(1-\gamma^{5}\right),\,2R=\left(1+\gamma^{5}\right)$ and
the coefficients $\varepsilon_{\alpha\beta}^{fP}$ encodes the deviation
from standard interactions between neutrinos of flavor $\alpha$
with component $P$-handed of fermions $f$, resulting in a neutrinos
of flavor $\beta$. Then, the neutrino oscillations in the presence
of non-standard matter effects can be described by an effective Hamiltonian, 
parameterized as \begin{equation}
\tilde{H}=\frac{1}{2E}\left[U\left(\begin{array}{ccc}
0 & 0 & 0\\
0 & \Delta m_{21}^{2} & 0\\
0 & 0 & \Delta m_{31}^{2}\end{array}\right)U^{\dagger}+a\left(\begin{array}{ccc}
1+\varepsilon_{ee} & \varepsilon_{e\mu} & \varepsilon_{e\tau}\\
\varepsilon_{e\mu}^{*} & \varepsilon_{\mu\mu} & \varepsilon_{\mu\tau}\\
\varepsilon_{e\tau}^{*} & \varepsilon_{\mu\tau}^{*} & \varepsilon_{\tau\tau}\end{array}\right)\right],\label{eq:NSI3}
\end{equation}
where $a=2\sqrt{2}G_{F}n_{f}$, $E$ is the neutrino energy 
and $\varepsilon_{\alpha\beta}=\sum_{f,P}\varepsilon_{\alpha\beta}^{fP}n_{f}/n_{e}$
with $n_{e}$ and $n_{f}$ the electrons and fermions $f$ density
in the middle, respectively. This parameters $\varepsilon_{\alpha\beta}$
can be found in solar\cite{nsis-1}, atmospheric\cite{nsia-1}, accelerator\cite{nsir-1}
and cross section\cite{nsic-1,nsic-2} neutrino data experiment. 

We focus on cross section neutrino experiment, where at low energies the
standard differential cross section for $\nu_{\alpha}e\rightarrow\nu_{\alpha}e$~
scattering processes has the well know form:\begin{equation}
\frac{d\sigma_{\alpha}}{dT}=\frac{2G_{F}m_{e}}{\pi}\left[\left(g_{1}^{\alpha}\right)^{2}+\left(g_{2}^{\alpha}\right)^{2}\left(1-\frac{T}{E_{\nu}}\right)^{2}-g_{1}^{\alpha}g_{2}^{\alpha}\frac{m_{e}T}{E_{\nu}^{2}}\right],\label{eq:choque8} \end{equation}
where $m_{e}$ is the electron mass, $E_{\nu}$ is the incident neutrino
energy, $T_{e}$ is the electron recoil energy. The quantities $g_{1}^{\alpha}$
and $g_{2}^{\alpha}$ are related to the SM neutral current couplings
of the electron $g_{L}^{e}=-1/2+\sin^{2}\theta_{W}$ and ~$g_{R}^{e}=\sin^{2}\theta_{W}$,
with $\sin^{2}\theta_{W}=0,23119$. For $\nu_{\mu,\tau}$ neutrinos, which
take part only in neutral current interactions, we have $g_{1}^{\mu,\tau}=g_{L}^{e}$
and $g_{2}^{\mu,\tau}=g_{R}^{e}$ while for electron neutrinos, 
which take part in  both
charge current (CC) and neutral current (NC) interactions, 
$g_{1}^{e}=1+g_{L}^{e},$ $g_{2}^{e}=g_{R}^{e}$.
In the presence of Non-Universal standard interaction the cross section
can be written in the same form of eq. (\ref{eq:choque8}) but with
$g_{1,2}^{\alpha}$ replaced by the effective non-standard couplings
$\tilde{g}_{1}^{\alpha}=g_{1}^{\alpha}+\varepsilon_{\alpha\alpha}^{eL}$
and $\tilde{g}_{2}^{\alpha}=g_{2}^{\alpha}+\varepsilon_{\alpha\alpha}^{eR}$,
leading to the following differential scattering cross section\cite{nsic-1,nsic-2}
\begin{eqnarray}
\frac{d\sigma_{\alpha}}{dT} & = & \frac{2G_{F}m_{e}}{\pi}\left\{ \left(g_{1}^{\alpha}+\varepsilon_{\alpha\alpha}^{eL}\right)^{2}+\left(g_{2}^{\alpha}+\varepsilon_{\alpha\alpha}^{eR}\right)^{2}\left(1-\frac{T_{e}}{E_{\nu}}\right)^{2}\right.\nonumber \\
 &  & \left.-\left(g_{1}^{\alpha}+\varepsilon_{\alpha\alpha}^{eL}\right)\left(g_{2}^{\alpha}+\varepsilon_{\alpha\alpha}^{eR}\right)\frac{m_{e}T_{e}}{E_{\nu}}\,\right\}. \label{eq:choque10}\end{eqnarray}

Our goal is to investigate how NSI with matter can be induced by new
physics generated by 331 models and based in constraint on this NSI
parameters constraint the model some values expected for 331 model. In 
section~\ref{sec:331model}

\section{331 Model}\label{sec:331model}

The success of the standard model (SM) implies  that any new theory should 
contain the symmetry $SU(3)_{C}\otimes SU(2)_{L}\otimes U(1)_{Y}$
($G_{321}$) in a low energy limit. Then, it is natural that one possible
modification of SM involves extensions of the representation content in
matter and Higgs sector, leading to extension of symmetry group $G_{321}$
to groups $SU(N_{C})_{C}\otimes SU(m)_{L}\otimes U(1)_{X}$ with $SU(N_{C})_{C}\otimes SU(m)_{L}\otimes U(1)_{X}\supset G_{321}$.

In early 90\textquoteright{}s, F. Pisano and V. Pleitez \cite{int-18,int-18-0}
and latter P. H. Frampton \cite{int-18-1} suggested an extension 
of the symmetry group $SU(2)_{L}\otimes U(1)_{Y}$ of electroweak sector
to a group $SU(3)_{L}\otimes U(1)_{X}$, {\it i.e.} with $N_{C}=m=3$. The 331
models present some interesting features, as for instance, they associate
the number of families to internal consistence of the theory, preserving
asymptotic freedom.

In these models, the SM doublets
are part of triplets. In quark sector three new quarks are included
to build the triplets, while in lepton sector we can use the right-handed
neutrino to such role\cite{int-18,int-18-1}. Another option is to
invoke three new heavy leptons, charged or not, depending on the choice
of charge operator\cite{int-18-2,int-18-3}. In SM the electric charge
operator is constructed as a combination of diagonal generators of
$SU(2)\otimes U(1)_{Y}$. Then, it is natural to assume that this
operator in $SU(3)_{L}\otimes U(1)_{X}$ is defined in the same way.
The most general charge operator in $SU(3)_{L}\otimes U(1)_{X}$ is
a linear superposition of diagonal generators of symmetry groups,
given by:\begin{equation}
\mathcal{Q}\equiv aT_{3L}+\frac{2}{\sqrt{3}}bT_{8L}+XI_{3},\label{eq:carga}\end{equation}
where the group generator are defined as $T_{iL}\equiv\lambda_{iL}/2$
with $\lambda_{iL}$, $i=1,...,8$ are Gell-Mann matrices for$SU(3)_{L}$,
where the normalization chosen is $Tr(\lambda_{iL}\lambda_{jL})=2\delta_{ij}$
and $I_{3}=\mbox{diag}\,(1,1,1)$ is the identity matrix, $a$ and $b$
are two parameters to be determined. Then, eq.~(\ref{eq:carga}) in
representation 3 have the form:
\begin{equation}
\mathcal{Q}[3]=\left(\begin{array}{ccc}
\frac{a}{2}+\frac{b}{3}+X & 0 & 0\\
0 & -\frac{a}{2}+\frac{b}{3}+X & 0\\
0 & 0 & -\frac{2b}{3}+X\end{array}\right),\label{eq:cargaMatriz}
\end{equation}
where we have two free parameters to obtain the charge of fermions, $a$ and $b$ 
($X$ can be determined by anomalies cancellation). However, $a=1$
is necessary to obtain doublets of isospins $SU(2)\otimes U(1)_{Y}$
correctly incorporated in the model $SU(3)_{L}\otimes U(1)_{X}$\cite{int-18-2,int-18-3,int-18-4}.
Then we can vary $b$ to create different models in 331 context, being
a signature which differentiate such models. For $b=-3/2$,
we have the original 331 model\cite{int-18,int-18-0}.

To have local gauge invariance we have the following covariant derivative:
$D_{\mu}=\partial_{\mu}-i\frac{g}{2}\lambda_{\alpha}W_{\mu}^{\alpha}-ig_{x}XB_{\mu}$
and a total of 17 mediator bosons: one field $B_{\mu}$ associated
with $U(1)_{X}$, eight fields associated with $SU(3)_{C}$, 
and another eight fields associated with $SU(3)_{L}$,  
written in the form:\begin{eqnarray}
\mathbf{W}_{\mu}\equiv W_{\mu}^{\alpha}\lambda_{\alpha} & = & \left(\begin{array}{ccc}
W_{\mu}^{3}+\frac{1}{\sqrt{3}}W_{\mu}^{8} & \sqrt{2}W_{\mu}^{+} & \sqrt{2}K_{\mu}^{\mathcal{Q}_{1}}\\
\sqrt{2}W_{\mu}^{-} & -W_{\mu}^{3}+\frac{1}{\sqrt{3}}W_{\mu}^{8} & \sqrt{2}K_{\mu}^{\mathcal{Q}_{2}}\\
\sqrt{2}K_{\mu}^{-\mathcal{Q}_{1}} & \sqrt{2}K_{\mu}^{-\mathcal{Q}_{2}} & -\frac{2}{\sqrt{3}}W_{\mu}^{8}\end{array}\right)\label{eq:W_mu},\end{eqnarray}
where

\begin{eqnarray}
W_{\mu}^{\pm} & = & \frac{1}{\sqrt{2}}\left(W_{1\mu}\mp iW_{2\mu}\right),\label{eq:22-4}\\
K_{\mu}^{\pm\mathcal{Q}_{1}} & = & \frac{1}{\sqrt{2}}\left(W_{4\mu}\mp iW_{5\mu}\right),\label{eq:22-3}\\
K_{\mu}^{\pm\mathcal{Q}_{2}} & = & \frac{1}{\sqrt{2}}\left(W_{6\mu}\mp iW_{7\mu}\right),\label{eq:22-2}\end{eqnarray}
Therefore, charge operator in eq. (\ref{eq:cargaMatriz}) applied over
eq. (\ref{eq:W_mu}) leads to:
\begin{equation}
\mathcal{Q}_{\mathbf{W}}\rightarrow\left(\begin{array}{ccc}
0 & +1 & \frac{1}{2}+b\\
-1 & 0 & -\frac{1}{2}+b\\
\frac{1}{2}+b & -\frac{1}{2}+b & 0\end{array}\right).\label{eq:Q_W}
\end{equation}
The mediator bosons will have integer electric charge only if $b=\pm1/2,\pm3/2,\pm5/2,...,\pm(2n+1)/2,n=0,1,2,3,...$
A detailed analysis shows that if $b$ is associated with the fundamental
representation 3 then $-b$ will be associated
with antisymmetric representation $3^{*}$.

\subsection{The representation content}

There are many representations for the matter content\cite{int-18-5},
for instance $b=3/2$ \cite{int-18}. But we note that if we accommodate
the doublets of $SU(2)_{L}$ in the superior components of triplets
and anti-triplets of $SU(3)_{L}$, and if we forbid exotic charges
for the new fermions, we obtain from eq. (\ref{eq:cargaMatriz}) the
constrain $b=\pm1/2$ (assuming $a=1$). Since a negative value of
$b$ can be associated to the anti-triplet, we obtain that $b=1/2$
is a necessary and sufficient condition to exclude exotic electric
charges in fermion and boson sector\cite{int-18-2}. 

The fields left and right-handed components transform under $SU(3)_{L}$
as triplets and singlets, respectively. Therefore the theory is quiral
and can present anomalies of Alder-Bell-Jackiw\cite{int-20,int-21,int-22-0}.
In a non-abelian theory, in the fermionic representation 
$\mathcal{\mathbf{\mathcal{R}}}$
the divergent anomaly is given by:
\begin{equation}
\mathcal{A}^{abc}\propto\sum_{\mathcal{R}}Tr\left[\{T_{L}^{a}(\mathcal{R}),T_{L}^{b}(\mathcal{R})\}T_{L}^{c}(\mathcal{R})-\{T_{R}^{a}(\mathcal{R}),T_{R}^{b}(\mathcal{R})\}T_{R}^{c}(\mathcal{R})\right],\label{eq:anomalia}
\end{equation}
where $T^{a}(\mathcal{R})$ are symmetry group matricial representation.
The indexes $R$ and $L$ relate to the quiral property of the fields.
Therefore, to eliminate the pure anomaly $[SU(3)_{L}]^{3}$ we should
have that $\mathcal{A}^{abc}\propto\sum_{\mathcal{R}'}Tr\left[\{T_{L}^{a}(\mathcal{R}'),T_{L}^{b}(\mathcal{R}')\}T_{L}^{c}(\mathcal{R}')\right]=0$.
We use the fact that $SU(3)_{L}$ has two fundamental representations,
$3$ and $3^{*}$, where $T^{a*}=-T^{a}$, which
is equivalent to say that $T_{L}^{a*}(\mathcal{R}^{*})=-T_{L}^{a}(\mathcal{R})$\cite{int-22-1}.
Then:\begin{eqnarray}
 &  & \sum_{\mathcal{R}'}Tr\left[\{T_{L}^{a}(\mathcal{R}'),T_{L}^{b}(\mathcal{R}')\}T_{L}^{c}(\mathcal{R}')\right]\nonumber \\
 &  & \,\,\,\,\,\,\,=\sum_{\mathcal{R}}Tr\left[\{T_{L}^{a}(\mathcal{R}),T_{L}^{b}(\mathcal{R})\}T_{L}^{c}(\mathcal{R})\right]+\sum_{\mathcal{R}^{*}}Tr\left[\{T_{L}^{a*}(\mathcal{R}^{*}),T_{L}^{b*}(\mathcal{R}^{*})\}T_{L}^{c*}(\mathcal{R}^{*})\right]\,\,\,\,\,\,\,\,\, \label{eq:anomali}\\
 &  & \,\,\,\,\,\,\,=\sum_{\mathcal{R}}Tr\left[\{T_{L}^{a}(\mathcal{R}),T_{L}^{b}(\mathcal{R})\}T_{L}^{c}(\mathcal{R})\right]-\sum_{\mathcal{R}}Tr\left[\{T_{L}^{a}(\mathcal{R}),T_{L}^{b}(\mathcal{R})\}T_{L}^{c}(\mathcal{R})\right].\label{eq:tripletosantitripletos}\end{eqnarray}
So, we can see that for the anomalies to be canceled, the number of fields
that transform as triplets (first term in eq. (\ref{eq:tripletosantitripletos})) and anti-triplets  under $SU(3)_{L}$ has to be the same. This implies that two families of
quarks should transform different then the third family, as will be
discussed in next section.

Usually the third quark family is chosed to transform in a different
way that the first two families. But we will assume that the first
family transform differently, to address the fact that $m_{u}<m_{d}$,
$m_{\nu_{\ell}}<m_{\ell}$ while $m_{c}>>m_{s}$ and $m_{t}>>m_{b}$.
To state in a more clear way, we remember that in SM the $SU(2)_{L}$
doublets are:$(\nu_{\ell},\ell)^{T}$, $(u,d)^{T}$, $(c,s)^{T}$, $(t,b)^{T}$, 
with $\ell=e,\mu,\tau$, We can see that the first component of
leptons doublets and first quark family is lighter that the second
component. But for the second and third quark families is the opposite.
Then we use this idea to justify that first quark family transform
as leptons.

\subsection{Minimal 331 model on scalar sector}

Among the different possibilities of 331 models, we will present a
detailed study on a Minimal Model on scalar sector without exotic
electric charges for quarks and with three news leptons without charged
\cite{int-18-2} $(b=1/2)$, where the fermions present the following
transformation structure under $SU(3)_{C}\otimes SU(3)_{L}\otimes U(1)_{X}$:
\begin{eqnarray}
\psi_{\ell L} & = & (\ell^{-},\,\nu_{\ell},\, N_{\ell}^{0})_{L}^{T}\sim(1,3^{*},-1/3),\nonumber \\
\nu_{\ell R} & \sim & (1,1,0),\nonumber \\
\ell_{R}^{-} & \sim & (1,1,-1),\nonumber \\
N_{\ell R}^{0} & \sim & (1,1,0),\nonumber \\
Q_{1L} & = & (d,\, u,U_{1})_{L}^{T}\sim(3,3^{*},1/3),\nonumber \\
u_{iR} & \sim & (3,1,2/3)\label{eq:modelo1},\\
d_{iR} & \sim & (3,1,-1/3),\nonumber \\
U_{1R} & \sim & (3,1,2/3),\nonumber \\
Q_{aL} & = & (u_{a},\, d_{a},\, D_{a})_{L}^{T}\sim(3,3,0),\nonumber \\
D_{aR} & \sim & (3,1,-1/3),\nonumber 
\end{eqnarray}
where $i=1,2,3$ , $\ell=e,\mu,\tau$, $a=2,3$. We note that the
leptons multiplets $\psi_{\ell L}$ consist of three fields $\ell=\{e,\mu,\tau\}$,
the corresponding neutrinos $\nu_{\ell}=\{\nu_{e},\nu_{\mu},\nu_{\tau}\}$
and new neutral leptons $N_{\ell}^{0}=\{N_{e}^{0},N_{\mu}^{0},N_{\tau}^{0}\}$.
We can also see that the multiplet associated with the first quark
family $Q_{1L}$
consists of quarks down, up and a new quark with the same electric
charge of quark up (named $U_{1}$),
while the multiplet associated with second (third) family $Q_{aL}$
consist of SM quarks of second (third) family and a new quark with
the same electric charge of quark down (named $D_{2}$ ($D_{3}$)).
The numbers on parenthesis refer to the transformation properties
under $SU(3)_{C}$ , $SU(3)_{L}$ and $U(1)_{X}$
respectively. With this choice the anomalies cancel in a non-trivial
way\cite{int-22}, and asymptotic freedom is guaranteed\cite{int-23,int-23-1}.

\subsubsection{Scalar sector and Yukawa couplings}

The scalar fields have to be coupled to fermions by Yukawa terms, invariants
under $SU(3)_{L}\otimes U(1)_{X}$. In lepton sector, these couplings
can be written as: 
\begin{eqnarray}
\bar{\psi}_{\ell L}\ell_{R} & \sim & (1,3,1/3)\otimes(1,1,-1)=\underbrace{(1,3,-2/3)}_{\rho^{*}},\nonumber \\
\bar{\psi}_{\ell L}\nu_{\ell R} & \sim & (1,3,1/3)\otimes(1,1,0)=\underbrace{(1,3,1/3)}_{\eta},\label{eq:leptons}\\
\bar{\psi}_{\ell L}N_{\ell R}^{0} & \sim & (1,3,1/3)\otimes(1,1,0)=\underbrace{(1,3,1/3)}_{\chi},\nonumber 
\end{eqnarray}
and in quarks sector:
\begin{eqnarray}
\bar{Q}_{1L}u_{iR} & = & (3^{*},3,-1/3)\otimes(3,1,2/3)=\underbrace{(1,3,1/3)}_{\chi}\oplus\underbrace{(8,3,1/3)}_{Colour\, Higgs},\nonumber \\
\bar{Q}_{1L}d_{iR} & = & (3^{*},3,-1/3)\otimes(3,1,-1/3)=\underbrace{(1,3,-2/3)}_{\rho^{*}}\oplus...,\nonumber \\
\bar{Q}_{1L}U_{1R} & = & (3^{*},3,-1/3)\otimes(3,1,2/3)=\underbrace{(1,3,1/3)}_{\chi}\oplus...,\nonumber \\
\bar{Q}_{aL}u_{iR} & = & (3^{*},3^{*},0)\otimes(3,1,2/3)=\underbrace{(1,3^{*},2/3)}_{\rho}\oplus...,\label{eq:quarks}\\
\bar{Q}_{aL}d_{iR} & = & (3^{*},3^{*},0)\otimes(3,1,-1/3)=\underbrace{(1,3^{*},-1/3)}_{\eta^{*}}\oplus...,\nonumber \\
\bar{Q}_{aL}D_{aR} & = & (3^{*},3^{*},0)\otimes(3,1,-1/3)=\underbrace{(1,3^{*},-1/3)}_{\chi^{*}}\oplus...,\nonumber 
\end{eqnarray}
As usual in these class of models, we impose non-colored Higgs,
selecting only the multiplets that transform as singlets under
$SU(3)_{C}$. We note that we need only three Higgs multiplets $\rho,\,\chi$
and $\eta,$ to couple the different fermionic fields and generate
mass through spontaneous symmetry breaking. In eqs. (\ref{eq:leptons})
and (\ref{eq:quarks}) we note that quantum numbers of triplets $\chi$
and $\eta$ are the same, which leads us to consider models with two or
three Higgs triplets. We will adopt the first option, two Higgs triplets,
due to the simpler scalar sector in comparison with the scenario with
three triplets\cite{int-18-2,int-18-3}.

\subsection{Model with two Higgs triplets}

For the models with two Higgs triplets, we obtain%
\footnote{Note that in this model we assumed $\Phi_{1}=\chi,\eta$ e $\Phi_{2}=\rho$%
}\begin{eqnarray}
\Phi_{1} & = & (\phi_{1}^{-},\phi_{1}^{'0},\phi_{1}^{0})^{T}\sim(1,3^{*},-1/3),\nonumber \\
\Phi_{2} & = & (\phi_{2}^{0},\phi_{2}^{+},\phi_{2}^{'+})^{T}\sim(1,3^{*},2/3).\label{eq:doshiggs}
\end{eqnarray}
Assuming the following
choice to the Higgs triplets vacuum expectation value (VEV) \cite{int-18-2}
$\langle\Phi_{1}\rangle_{0}=(0,\vartheta_{1},V)^{T}$ and $\langle\Phi_{2}\rangle_{0}=(\vartheta_{2},0,0)^{T}$
we associate $V$ with the mass of the new fermions, which lead us
to assume $V>>\vartheta_{1},\vartheta_{2}$. We expand the scalar
VEV\textquoteright{}s in the following way:\begin{equation}
\phi_{1}^{0}=V+\frac{H_{\phi_{1}}^{0}+iA_{\phi_{1}}^{0}}{\sqrt{2}},\,\,\,\,\phi_{1}^{'0}=\vartheta_{1}+\frac{H_{\phi_{1}}^{'0}+iA_{\phi_{1}}^{'0}}{\sqrt{2}},\,\,\,\,\phi_{2}^{0}=\vartheta_{2}+\frac{H_{\phi_{2}}^{0}+iA_{\phi_{2}}^{0}}{\sqrt{2}}.\label{eq:componesnteshiggs}\end{equation}
The real (imaginary) part $H_{\phi_{i}}$ ($A_{\phi_{i}}$ ) is usually
called CP-even (CP-odd) scalar field. The most general potential can
be written as:\begin{eqnarray}
V(\Phi_{1},\Phi_{2}) & = & \mu_{1}^{2}\Phi_{1}^{\dagger}\Phi_{1}+\mu_{2}^{2}\Phi_{2}^{\dagger}\Phi_{2}+\lambda_{1}\left(\Phi_{1}^{\dagger}\Phi_{1}\right)^{2}+\lambda_{2}\left(\Phi_{2}^{\dagger}\Phi_{2}\right)^{2}\nonumber \\
 &  & +\lambda_{3}\left(\Phi_{1}^{\dagger}\Phi_{1}\right)\left(\Phi_{2}^{\dagger}\Phi_{2}\right)+\lambda_{4}\left(\Phi_{1}^{\dagger}\Phi_{2}\right)\left(\Phi_{2}^{\dagger}\Phi_{1}\right)\label{eq:potencial}.\end{eqnarray}
Demanding that in the displaced potential $V(\Phi_{1},\Phi_{2})$ the
linear terms on the field should be absent, we have, in tree level
approximation, the following constraints:\begin{eqnarray}
\mu_{1}^{2}+2\lambda_{1}\left(\vartheta_{1}^{2}+V^{2}\right)+\lambda_{3}\vartheta_{2}^{2} & = & 0,\nonumber \\
\mu_{2}^{2}+\lambda_{3}\left(\vartheta_{1}^{2}+V^{2}\right)+2\lambda_{2}\vartheta_{2}^{2} & = & 0.\label{eq:restripotencial}\end{eqnarray}
The analysis of such equations shows that they are related to a minimum
in scalar potential with the value:\begin{equation}
V_{min}=-\vartheta_{2}^{4}\lambda_{2}-\left(\vartheta_{1}^{2}+V^{2}\right)\left[\left(\vartheta_{1}^{2}+V^{2}\right)\lambda_{1}+\vartheta_{2}^{2}\lambda_{3}\right]=V(\vartheta_{1},\vartheta_{2},V)\label{eq:potencialminimo}.\end{equation}
Replacing
eq. (\ref{eq:componesnteshiggs}) and (\ref{eq:restripotencial})
in eq. (\ref{eq:potencial}) we can calculate the mass matrix in $\left(H_{\phi_{1}}^{0},\, H_{\phi_{2}}^{0},\, H_{\phi_{1}}^{'0}\right)$
basis through the relation $M_{ij}^{2}=2\frac{\partial^{2}V\left(\Phi_{1},\Phi_{2}\right)}{\partial H_{\Phi_{i}}^{0}\partial H_{\Phi_{j}}^{0}}$
, obtaining:

\begin{equation}
M_{H}^{2}=2\left(\begin{array}{ccc}
2\lambda_{1}V^{2} & \lambda_{3}\vartheta_{2}V & \,2\lambda_{1}\vartheta_{1}V\\
\lambda_{3}\vartheta_{2}V & 2\lambda_{2}\vartheta_{2}^{2} & \lambda_{3}\vartheta_{1}\vartheta_{2}\\
2\lambda_{1}\vartheta_{1}V & \lambda_{3}\vartheta_{1}\vartheta_{2} & 2\lambda_{1}\vartheta_{1}^{2}\end{array}\right).\label{eq:massa}\end{equation}
Since eq. (\ref{eq:massa}) has vanishing determinant, we have one
Goldstone boson $G_{1}$ and two massive neutral scalar fields $H_{1}$
and $H_{2}$ with masses %
\footnote{Note that if $\lambda_{3}^{2}=4\lambda_{1}\lambda_{2}$ we obtain
two Goldstone bosons, $G_{1}$ and $H_{2}$, and a massive scalar
field $H_{1}$ with mass $M_{H_{1}}^{2}=4\left[\lambda_{1}\left(\vartheta_{1}^{2}+V^{2}\right)+\lambda_{2}\vartheta_{2}^{2}\right]$
where $\lambda_{1}\lambda_{2}>0$, then imposing $M_{H_{1}}^{2}>0$
leads to $\lambda_{1}>0$ and $\lambda_{2}>0$.%
}\begin{eqnarray}
M_{H_{1},H_{2}}^{2} & = & 2\lambda_{1}\left(\vartheta_{1}^{2}+V^{2}\right)+2\lambda_{2}\vartheta_{2}^{2}\nonumber \\
 & \pm & 2\sqrt{\left[\lambda_{1}\left(\vartheta_{1}^{2}+V^{2}\right)+\lambda_{2}\vartheta_{2}^{2}\right]^{2}+\vartheta_{2}^{2}\left(\vartheta_{1}^{2}+V^{2}\right)\left(\lambda_{3}^{2}-4\lambda_{1}\lambda_{2}\right)}\label{eq:masmassas}~~,\end{eqnarray}
where real values for $\lambda$'s produce positive mass to neutral
scalar fields only if $\lambda_{1}>0$ and $4\lambda_{1}\lambda_{2}>\lambda_{3}^{2}$,
which implies that $\lambda_{2}>0$. A detailed analysis shows that
when $V(\Phi_{1},\Phi_{2})$ in eq. (\ref{eq:potencial}) is expanded
around the most general vacuum, given by eq. (\ref{eq:componesnteshiggs})
and using constrains in eq. (\ref{eq:restripotencial}), we don\textquoteright{}t
obtain pseudo-scalar fields $A_{\Phi_{i}}^{0}$. This allows us do
identify three more Goldstone bosons $G_{2}=A_{\Phi_{1}}^{0}$, $G_{3}=A_{\Phi_{2}}^{0}$
and $G_{4}=A_{\Phi_{1}}^{'0}$ . For the mass spectrum in charged
scalar sector on $\left(\phi_{1}^{-},\,\phi_{2}^{+},\,\,\phi_{2}^{'+}\right)$
basis the mass matrix will be given by:\begin{equation}
M_{+}^{2}=2\lambda_{4}\left(\begin{array}{ccc}
\vartheta_{2}^{2} & \vartheta_{1}\vartheta_{2} & \vartheta_{2}V\\
\vartheta_{1}\vartheta_{2} & \vartheta_{1}^{2} & \vartheta_{1}V\\
\vartheta_{2}V & \vartheta_{1}V & V^{2}\end{array}\right),\label{eq:matrizmassas}\end{equation}
with two eigenvalues equal to zero, equivalent to four Goldstone bosons
$G_{5}^{\pm},\, G_{6}^{\pm}$ and two physical charged scalar fields
with large masses given by $\lambda_{4}\left(\vartheta_{1}^{2}+\vartheta_{2}^{2}+V^{2}\right),$
which leads to the constrain $\lambda_{4}>0$.

This analysis shows that, after symmetry breaking, the original twelve
degrees of freedom in scalar sector leads to eight Goldstone bosons
(four electrically neutral and four electrically charged), four physical
scalar fields, two neutral (one of which being the SM Higgs scalar)
and two charged. Eight Goldstone bosons should be absorbed by eight
gauge fields as we will see in next section.

\subsubsection{Gauge Sector with two Higgs triplets\label{sub:Gauge-Sector-with}}

The gauge bosons interaction with matter in electroweak sector appears
with the covariant derivative for a matter field 
$\varphi$ as:
\begin{eqnarray}
D_{\mu}^{\varphi} & = & \partial_{\mu}-\frac{i}{2}gW_{\mu}^{a}\lambda_{aL}-ig_{X}X_{\varphi}B_{\mu}=\partial_{\mu}-\frac{i}{2}g\mathcal{M}_{\mu}^{\varphi}\label{eq:20}
,\end{eqnarray}
where $\lambda_{aL},\, a=1,\,...,\,8$ are Gell-Mann matrices of $SU(3)_{L}$
algebra, and $X_{\varphi}$ is the charge of abelian factor $U(1)_{X}$
of the multiplet $\varphi$ in which $D_{\mu}$ acts . The matrix
$\mathcal{M}_{\mu}^{\varphi}$ contain the gauge bosons with electric
charges q, defined by the generic charge operator in eq. (\ref{eq:carga})
and eq. (\ref{eq:Q_W}). For $b=1/2$ the matrix $\mathcal{M}_{\mu}^{\varphi}$
will have the form:\begin{eqnarray}
\mathcal{M}_{\mu}^{\varphi} & = & \left(\begin{array}{ccc}
W_{3\mu}+\frac{W_{8\mu}}{\sqrt{3}}+2tX_{\varphi}B_{\mu} & \sqrt{2}W_{\mu}^{+} & \sqrt{2}K_{\mu}^{+}\\
\sqrt{2}W_{\mu}^{-} & -W_{3\mu}+\frac{W_{8\mu}}{\sqrt{3}}+2tX_{\varphi}B_{\mu} & \sqrt{2}K_{\mu}^{0}\\
\sqrt{2}K_{\mu}^{-} & \sqrt{2}\bar{K}_{\mu}^{0} & \frac{-2W_{8\mu}}{\sqrt{3}}+2tX_{\varphi}B_{\mu}\end{array}\right)\,\,\,\,\,\,\,\,\,\label{eq:21},\end{eqnarray}
where $t=g_{x}/g$ and non physical gauge bosons on non-diagonal entries,$W_{\mu}^{\pm}\,\,\mbox{and\,\,}K_{\mu}^{\pm}$,
are defined in eq. (\ref{eq:22-4}) and (\ref{eq:22-3}) with $\mathcal{Q}_{1}=1$
respectively, and:\begin{eqnarray}
K_{\mu}^{0} & = & \frac{1}{\sqrt{2}}\left(A_{6\mu}-iA_{7\mu}\right),\label{eq:22-1}\\
\bar{K}_{\mu}^{0} & = & \frac{1}{\sqrt{2}}\left(A_{6\mu}+iA_{7\mu}\right).\end{eqnarray}
Then for the 331 model we are considering ($b=1/2$) we have two neutral
gauge bosons, $K_{\mu}^{0}$ and $\bar{K}_{\mu}^{0}$, and four charged
gauge bosons, $W_{\mu}^{\pm}$ and $K_{\mu}^{\pm}$. The three physical
neutral eigenstates will be a linear combination of $W_{3\mu},\, W_{8\mu}$
and $B_{\mu}.$ After breaking the symmetry with $\langle\Phi_{i}\rangle$,
$i=1,2,$ and using covariant derivative $D_{\mu}=\partial_{\mu}-\frac{i}{2}g\mathcal{M}_{\mu}^{\varphi}$
for the triplets $\Phi_{i}$ we obtain the following masses for the
charged physical fields:\begin{equation}
M_{W'}^{2}=\frac{1}{2}g^{2}\vartheta_{2}^{2},\,\,\,\, M_{K'}^{2}=\frac{1}{2}g^{2}\left(\vartheta_{1}^{2}+\vartheta_{2}^{2}+V^{2}\right),\label{eq:24}\end{equation}
and the following physical eigenstates: \begin{equation}
W_{\mu}^{'\pm}=\frac{1}{\sqrt{\vartheta_{1}^{2}+V^{2}}}\left(-\vartheta_{1}K_{\mu}^{\pm}+VW_{\mu}^{\pm}\right),\,\,\, K_{\mu}^{'\pm}=\frac{1}{\sqrt{\vartheta_{1}^{2}+V^{2}}}\left(VK_{\mu}^{\pm}+\vartheta_{1}W_{\mu}^{\pm}\right).\label{eq:25}\end{equation}
The neutral sector in approximation $\left(\frac{\vartheta_{i}}{V}\right)^{n}\approx0$
for $n>2$ leads to the following masses for the neutral physical
fields: \begin{eqnarray}
M_{f\acute{o}ton}^{2} & = & 0 ~,\nonumber \\
M_{Z'}^{2} & = & \frac{1}{2}g^{2}\left(V^{2}+\vartheta_{1}^{2}\right),\nonumber \\
M_{Z}^{2} & \approx & \frac{1}{2}g^{2}\vartheta_{2}^{2}\left(\frac{3g^{2}+4g_{x}^{2}}{3g^{2}+g_{x}^{2}}\right)\label{eq:28},\\
M_{K_{R}^{0}}^{2} & \approx & \frac{2}{9}\left(V^{2}+\vartheta_{1}^{2}\right)\left(3g^{2}+g_{x}^{2}\right)+\frac{\vartheta_{2}^{2}\left(3g^{2}+4g_{x}^{2}\right)^{2}}{18\left(3g^{2}+g_{x}^{2}\right)},\nonumber \\
M_{K_{I}^{0}}^{2} & = & \frac{1}{2}g^{2}\left(V^{2}+\vartheta_{1}^{2}\right).\nonumber \end{eqnarray}
We can see from eqs. (\ref{eq:24}) and (\ref{eq:28}) that we have
one non-massive boson, which we associate with the photon, and four massive
neutral fields, where the mass of one of them is proportional to $\vartheta_{2}$
while the other three have masses proportional to $V$ (new energy
scale). Therefore we can associate the field $Z$ with SM ${Z}_{\mu}$,
and the fields $Z'$, $K_{I}^{0}$ and $K_{R}^{0}$ , with three new
neutral bosons. We also have four massive charged fields, where two
of them have masses proportional to $\vartheta_{2}$. Therefore we
can associate the fields $W_{\mu}^{'\pm}$ to the SM fields ${W}_{\mu}^{\pm}$
, while the fields $K_{\mu}^{'\pm}$ are new bosons. The eigenstates
$B_{\mu},\, W_{3\mu},\, W_{8\mu}$ and $K_{R\mu}^{o}$ can be related
to the physical eigenstates $A_{\mu},\, Z_{\mu}^{'0},\, Z_{\mu}^{0}$
and $K_{{R}\mu}^{'0}$ by:\begin{equation}
\left(\begin{array}{c}
B_{\mu}\\
W_{3\mu}\\
W_{8\mu}\\
K_{R\mu}^{o}\end{array}\right)=\mathbf{M^{-1}}\left(\begin{array}{c}
A_{\mu}\\
Z_{\mu}^{'0}\\
Z_{\mu}^{0}\\
K_{{R}\mu}^{'0}\end{array}\right).\label{eq:29}\end{equation}
Assuming $\left(\frac{\vartheta_{i}}{V}\right)^{n}\sim0$ for $n>2$,
we obtain \begin{equation}
\mathbf{M}^{-1}=\left(\begin{array}{cccc}
-\frac{1}{t}S_{W} & 0 & \frac{1}{t}T_{W}^{2}C_{W}+\beta_{1} & -\frac{1}{\sqrt{3}}T_{W}+\beta_{2}\\
S_{W} & \frac{-\vartheta_{1}}{V} & C_{W}+\beta_{3} & \beta_{4}\\
\frac{1}{\sqrt{3}}S_{W} & \frac{\sqrt{3}\vartheta_{1}}{V} & -\frac{1}{\sqrt{3}}T_{W}S_{W}+\beta_{5} & -\frac{1}{t}T_{W}+\beta_{6}\\
0 & 1-\beta_{7} & \frac{\vartheta_{1}}{V}C_{W}^{-1} & \frac{\sqrt{3}\vartheta_{1}}{tV}T_{W}\end{array}\right)\label{eq:31},\end{equation}
where, again, $t=g_{x}/g$ and: \begin{eqnarray}
S_{W} & = & \frac{\sqrt{3}g_{x}}{\sqrt{3g^{2}+4g_{x}^{2}}},\,\, C_{W}=\sqrt{1-S_{W}^{2}},\,\, T_{W}=\frac{S_{W}}{C_{W}},\nonumber \\
\beta_{1} & = & -\frac{\vartheta_{2}^{2}}{4tV^{2}}T_{W}^{2}C_{W}^{-3},\,\,\,\,\, \beta_{2}=-\frac{\sqrt{3}\vartheta_{2}^{2}}{4t^{2}V^{2}}T_{W}^{3}C_{W}^{-2},\nonumber \\
\beta_{3} & = & -\frac{\vartheta_{1}^{2}}{2V^{2}}C_{W}^{-1},\,\,\,\,\,\beta_{4}=-\frac{\sqrt{3}\left(2C_{W}^{2}\vartheta_{1}^{2}+\vartheta_{2}^{2}\right)}{4tV^{2}}T_{W}C_{W}^{-2},\nonumber \\
\beta_{5} & = & \frac{6C_{W}^{4}\vartheta_{1}^{2}-\left(3-4S_{W}^{2}\right)\vartheta_{2}^{2}}{4\sqrt{3}V^{2}C_{W}^{5}},\,\, \beta_{6}=\frac{\left(6C_{W}^{4}\vartheta_{1}^{2}+S_{W}^{2}\vartheta_{2}^{2}\right)}{4tV^{2}C_{W}^{4}}T_{W},\nonumber \\
\beta_{7} & = & -\frac{2\vartheta_{2}^{2}}{V^{2}}.\label{eq:-6}\end{eqnarray}
We note that all $\beta_{i}$ are of order $\mathcal{O}\left(\left(\frac{\vartheta_{i}}{V}\right)^{2}\right)$.
So, assuming $\vartheta_{i}\sim\mathcal{O}\left(10^{-1}\right)\,\mathrm{TeV}$,
for a new energy scale of order $V\sim10\,\mathrm{TeV}$ all the $\beta_{i}'$s
are negligible.

\subsubsection{Charged and Neutral Currents}

The interaction between gauge bosons and fermions in flavor basis
is given by the following Lagrangian density:

\begin{equation}
\mathcal{L}_{f}=\bar{R}i\gamma^{\mu}(\partial_{\mu}+ig_{x}B_{\mu}X_{R})R+\bar{L}i\gamma^{\mu}(\partial_{\mu}+ig_{x}B_{\mu}X_{L}+\frac{ig}{2}\lambda_{a}W_{\mu}^{a})L\label{eq:0},\end{equation}
where $R$ represents any right-handed singlet and $L$ any left-handed
triplet. We can write $\mathcal{L}_{f}=\mathcal{L}_{lep}+\mathcal{L}_{Q_{1}}+\mathcal{L}_{Q_{a}}$
and in lepton sector we obtain:\begin{eqnarray}
\mathcal{L}_{lep} & = & \mathcal{L}_{lep}^{kin}+\mathcal{L}_{lep}^{CC}+\mathcal{L}_{lep}^{NC}\label{eq:3},\end{eqnarray}
where

\begin{eqnarray}
\mathcal{L}_{lep}^{kin} & = & \bar{R}i\gamma^{\mu}\partial_{\mu}R+\bar{L}i\gamma^{\mu}\partial_{\mu}L\label{eq:4},\\
\mathcal{L}_{lep}^{CC} & = & -\frac{g}{\sqrt{2}}\bar{\ell_{L}}\gamma^{\mu}\nu_{\ell L}W_{\mu}^{+}-\frac{g}{\sqrt{2}}\bar{\ell_{L}}\gamma^{\mu}N_{\ell L}^{0}K_{\mu}^{+}+\,\, h.c.,\label{eq:5}\\
\mathcal{L}_{lep}^{NC} & = & \frac{g_{x}}{3}\left[\bar{\ell_{L}}\gamma^{\mu}\ell+\overline{\nu_{\ell L}}\gamma^{\mu}\nu_{\ell L}+\overline{N_{\ell L}^{0}}\gamma^{\mu}N_{\ell L}^{0}\right]B_{\mu}+g_{x}\overline{\ell_{R}}\gamma^{\mu}\ell_{R}B_{\mu}\nonumber \\
 & - & \frac{g}{2\sqrt{3}}\left[\overline{\ell_{L}}\gamma^{\mu}\ell_{L}+\overline{\nu_{\ell L}}\gamma^{\mu}\nu_{\ell L}-2t\overline{N_{\ell L}^{0}}\gamma^{\mu}N_{\ell L}^{0}\right]W_{8\mu}-\frac{g}{\sqrt{2}}\overline{\nu_{\ell L}}\gamma^{\mu}N_{\ell L}^{0}K^{0\mu}\nonumber \\
 & - & \frac{g}{2}\left[\overline{\ell_{L}}\gamma^{\mu}\ell_{L}-\overline{\nu_{\ell L}}\gamma^{\mu}\nu_{\ell L}\right]W_{3\mu}-\frac{g}{\sqrt{2}}\overline{N_{\ell L}^{0}}\gamma^{\mu}\nu_{\ell L}\bar{K}_{\mu}^{0}\label{eq:6}.\end{eqnarray}
In quark sector we have that for the first family triplet $X=1/3$,
and for the singlets $d,\, u$, and $U_{1}$ we have $X=-1/3,\,2/3$
and $2/3$, respectively. Then we have\begin{eqnarray}
\mathcal{L}_{Q_{1}}^{kin} & = & \bar{Q}_{1R}i\gamma^{\mu}\partial_{\mu}Q_{1R}+\bar{Q}_{1L}i\gamma^{\mu}\partial_{\mu}Q_{1L},\label{eq:14}\\
\mathcal{L}_{Q_{1}}^{CC} & =- & \frac{g}{\sqrt{2}}\overline{d_{L}}\gamma^{\mu}u_{L}W_{\mu}^{+}-\frac{g}{\sqrt{2}}\overline{d_{L}}\gamma^{\mu}U_{1L}K_{\mu}^{+}+\, h.c.,\label{eq:15}\\
\mathcal{L}_{Q_{1}}^{NC} & = & \frac{g_{x}}{3}\left(\overline{d_{R}}\gamma^{\mu}d_{R}-2\overline{u_{R}}\gamma^{\mu}u_{R}-2\overline{U_{1R}}\gamma^{\mu}U_{1R}\right)B_{\mu}+\frac{g}{2}\overline{u_{L}}\gamma^{\mu}u_{L}W_{3\mu}\nonumber \\
 & - & \frac{g_{x}}{3}\left(\overline{d_{L}}\gamma^{\mu}d_{L}+\overline{u_{L}}\gamma^{\mu}u_{L}+\overline{U_{1L}}\gamma^{\mu}U_{1L}\right)B_{\mu}-\frac{g}{2}\overline{d_{L}}\gamma^{\mu}d_{L}W_{3\mu}-\frac{g}{\sqrt{2}}\overline{U_{1L}}\gamma^{\mu}u_{L}\bar{K}_{\mu}^{0}\nonumber \\
 & - & \frac{g}{2\sqrt{3}}\left(\overline{d_{L}}\gamma^{\mu}d_{L}+\overline{u_{L}}\gamma^{\mu}u_{L}-2\overline{U_{1L}}\gamma^{\mu}U_{1L}\right)W_{8\mu}-\frac{g}{\sqrt{2}}\overline{u_{L}}\gamma^{\mu}U_{1L}K_{\mu}^{0}\label{eq:16}.\end{eqnarray}

For second and third families we know that $X=0$ for the triplets
and $X=2/3,\,-1/3$ and $-1/3$
for the singlets $u_{2,3},\, d_{2,3},\, D_{2,3}$, respectively, where
$u_{2}=c$, $u_{3}=t$, $d_{2}=s$,$d_{3}=b$. Then we obtain for
$a=2,\,3$:\begin{eqnarray}
\mathcal{L}_{Q_{a}}^{kin} & = & \bar{Q}_{aR}i\gamma^{\mu}\partial_{\mu}Q_{aR}+\bar{Q}_{aL}i\gamma^{\mu}\partial_{\mu}Q_{aL},\label{eq:17}\\
\mathcal{L}_{Q_{a}}^{CC} & = & -\frac{g}{\sqrt{2}}\overline{u_{aL}}\gamma^{\mu}d_{aL}W_{\mu}^{+}-\frac{g}{\sqrt{2}}\overline{u_{aL}}\gamma^{\mu}D_{aL}K_{\mu}^{+}+\,\, h.c.,\label{eq:18}\\
\mathcal{L}_{Q_{a}}^{NC} & = & \frac{g_{x}}{3}\left[-2\overline{u_{aR}}\gamma^{\mu}u_{aR}+\overline{d_{aR}}\gamma^{\mu}d_{aR}+\overline{D_{aR}}\gamma^{\mu}D_{aR}\right]B_{\mu}\nonumber \\
 & - & \frac{g}{2\sqrt{3}}\left[\overline{u_{aL}}\gamma^{\mu}u_{aL}+\overline{d_{aL}}\gamma^{\mu}d_{aL}-4\overline{D_{aL}}\gamma^{\mu}D_{aL}\right]W_{8\mu}-\frac{g}{\sqrt{2}}\overline{d_{aL}}\gamma^{\mu}D_{aL}K_{\mu}^{0}\nonumber \\
 & - & \frac{g}{2}\left[\overline{u_{aL}}\gamma^{\mu}u_{aL}-\overline{d_{aL}}\gamma^{\mu}d_{aL}\right]W_{3\mu}-\frac{g}{\sqrt{2}}\overline{D_{aL}}\gamma^{\mu}d_{aL}\bar{K}_{\mu}^{0}.\label{eq:19}\end{eqnarray}

\section{NEUTRINOS INTERACTIONS WITH MATTER IN 331 MODEL }

It is well known that neutrino oscillation phenomenon in a material
medium, as the sun, earth or in a supernova, can be quite different
from the oscillation that occurs in vacuum, since the interactions
in the medium modify the dispersion relations of the particles traveling
through it \cite{int-24}. From the macroscopic point of view, the
modifications of neutrino dispersion relations can be represented
in terms of a refractive index or an effective potential. And according
to \cite{int-24,int-25}, the effective potential can be calculated
from the amplitudes of coherent elastic scattering in relativistic
limit.

In the present 331 model, the coherent scattering will be induced
by neutral currents, NC, mediated by bosons $Z_{\mu}^{'0},\, Z_{\mu}^{0},\,\,\mbox{and\,}\, K_{{R}\mu}^{'0}$
and by charged currents, CC, mediated by bosons $W_{\mu}^{'\pm}$
and $K_{\mu}^{'\pm}$. Following \cite{int-25}, we calculate in next
sections the neutrino effective potentials in coherent scattering.

\subsection{Charged Currents\label{sub:Charged-Currents}}

The first term of eq. (\ref{eq:5}) shows that the interaction of
charged leptons with neutrinos occurs only through the gauge bosons
$W_{\mu}^{\pm}$ , then by eq. (\ref{eq:25}) we obtain that the interaction
through charged bosons is given by:\begin{eqnarray}
-\frac{g}{\sqrt{2}}\overline{\ell_{L}}\gamma^{\mu}\nu_{\ell L}W_{\mu}^{+} & = & -\frac{Vg}{\sqrt{2}\sqrt{\vartheta_{1}^{2}+V^{2}}}\bar{\ell_{L}}\gamma^{\mu}\nu_{\ell L}W_{\mu}^{'\pm}-\frac{g\vartheta_{1}}{\sqrt{2}\sqrt{\vartheta_{1}^{2}+V^{2}}}\bar{\ell_{L}}\gamma^{\mu}\nu_{\ell L}K_{\mu}^{'\pm}\label{eq:3.1}.\end{eqnarray}
The amplitude for the neutrino elastic scattering with charged leptons
in tree level through CC is given by %
\footnote{Note from eq. (\ref{eq:3.1}) that only left-handed
leptons interact with neutrinos, as in SM.%
}\begin{eqnarray}
\mathcal{L}_{int}^{cc} & = & -\left(-\frac{Vg}{\sqrt{2}\sqrt{\vartheta_{1}^{2}+V^{2}}}\right)^{2}\bar{\ell}_{L}(p_{1})\gamma^{\mu}\nu_{\ell L}(p_{2})\frac{-ig_{\mu\lambda}}{(p_{2}-p_{1})^{2}-M_{W'}^{2}}\bar{\nu}_{\ell L}(p_{3})\gamma^{\lambda}\ell_{L}(p_{4})\nonumber \\
 & - & \left(-\frac{g\vartheta_{1}}{\sqrt{2}\sqrt{\vartheta_{1}^{2}+V^{2}}}\right)^{2}\bar{\ell}_{L}(p_{1})\gamma^{\mu}\nu_{\ell L}(p_{2})\frac{-ig_{\mu\lambda}}{(p_{2}-p_{1})^{2}-M_{K'}^{2}}\bar{\nu}_{\ell L}(p_{3})\gamma^{\lambda}\ell_{L}(p_{4})\label{eq:3.2}.\end{eqnarray}
For low energies $M_{W'}^{2},\, M_{K'}^{2}>>(p_{2}-p_{1})^{2}$, the
effective Lagrangian is given by:\begin{eqnarray}
\mathcal{L}_{eff}^{cc} & \approx- & \frac{g^{2}}{2\left(\vartheta_{1}^{2}+V^{2}\right)}\left(\frac{V^{2}}{M_{W'}^{2}}+\frac{\vartheta_{1}^{2}}{M_{K'}^{2}}\right)\left[\bar{\ell}_{L}(p_{1})\gamma^{\mu}\ell_{L}(p_{4})\right]\left[\bar{\nu}_{\ell L}(p_{3})\gamma_{\mu}\nu_{\ell L}(p_{2})\right],\label{eq:3.3}\end{eqnarray}
where we used Fierz transformation\cite{int-30} to go from eq. (\ref{eq:3.2})
to eq. (\ref{eq:3.3}). Replacing eq. (\ref{eq:24}) in eq. (\ref{eq:3.3})
we obtain:
\begin{eqnarray}
-\mathcal{L}_{eff}^{cc} & \approx & \left[\frac{1}{\vartheta_{2}^{2}}-\frac{\vartheta_{1}^{2}}{V^{2}\vartheta_{2}^{2}}+\left(\frac{\vartheta_{1}^{2}}{V^{4}}\right)_{K'}+\mathcal{O}\left(\frac{1}{V^{4}}\right)\right]\left\langle \bar{\ell}\gamma^{\mu}\frac{(1-\gamma_{5})}{2}\ell\right\rangle \left\{ \bar{\nu}_{\ell L}(p)\gamma^{\mu}\nu_{\ell L}(p)\right\}, \label{eq:3.4}
\end{eqnarray}
where we used
$\left(\right)_{K'}$ to denote the term that appears from
the new charged boson. We can see that for a new energy scale 
$V\gg\vartheta_{1}$
the term that comes from the new boson does not contribute to the
process, as expected, since the new charged boson $K_{\mu}^{'\pm}$
has a mass of the order of the new energy scale of the theory (see
eq.~(\ref{eq:24})).

Now, since usual matter has only leptons from first family, we will
restrain our calculations to the neutrino interactions with first
family standard model particles. The term $\langle\,\rangle$ in eq.
(\ref{eq:3.4}) can be calculated following \cite{int-25}, where we
have the correspondence $\left\langle \bar{e}\gamma^{\mu}\gamma_{5}e\right\rangle \sim$ spin,
$\left\langle \bar{e}\gamma_{i}e\right\rangle \sim$ velocity
and $\left\langle \bar{e}\gamma_{0}e\right\rangle \sim n_{e}$, where
$n_{e}$ is the electronic density. Assuming
non-polarized medium and vanishing average velocity, we obtain that
eq. (\ref{eq:3.4}) can be written as:\begin{equation}
\mathcal{L}_{eff}^{cc}\approx-\left[\frac{1}{2\vartheta_{2}^{2}}-\frac{\vartheta_{1}^{2}}{2V^{2}\vartheta_{2}^{2}}+\left(\frac{\vartheta_{1}^{2}}{2V^{4}}\right)_{K'}+\mathcal{O}\left(V^{-4}\right)\right]n_{e}\bar{\nu}_{eL}\gamma^{\mu}\nu_{eL}.\label{eq:3.4-1}\end{equation}
The modifications on electronic neutrino dispersion relations can
be represented by the following effective potential:\begin{equation}
V_{CC}^{e}\approx\frac{1}{2\vartheta_{2}^{2}}n_{e}-\frac{\vartheta_{1}^{2}}{2V^{2}\vartheta_{2}^{2}}n_{e}+\left(\frac{\vartheta_{1}^{2}}{2V^{4}}\right)_{K'}n_{e}+\mathcal{O}\left(V^{-4}\right).\label{eq:3.5}\end{equation}
Disregarding the term $()_{K'}$ since we are assuming $V\gg\vartheta_{i}$,
and remembering that in subsection \ref{sub:Gauge-Sector-with} we
associated boson $W'$ with SM boson ${W}$, we can easily
associate:\begin{equation}
\sqrt{2}G_{F}\approx\frac{1}{2\vartheta_{2}^{2}}-\frac{\vartheta_{1}^{2}}{2V^{2}\vartheta_{2}^{2}}.\label{eq:3.6}\end{equation}
We note that eq. (\ref{eq:3.6}) gives limits for the VEV of one of
Higgs triplets. Under assumption $\vartheta_{1}\sim\vartheta_{2}\ll V$, 
we can write $G_{F}\approx\frac{1}{2\sqrt{2}\vartheta_{2}^{2}}\left(1-\frac{\vartheta_{1}^{2}}{V^{2}}\right)$,
from which can see that the maximum value of $\vartheta_{2}^{2}$
is achieved when we consider $\frac{\vartheta_{1}^{2}}{V^{2}}=0$,
in which replacing $G_{F}=1.166\,37(1)\times10^{-5}\mathrm{Gev}^{-2}$
leads to 
\begin{equation}
\vartheta_{2}\lesssim174.105\,\mathrm{GeV}~~~.
\end{equation}

\subsection{Neutral Current}

The lagrangian for neutrino elastic-scattering with fermions $f=e,\, u,\, d$
through NC is given by:\begin{eqnarray}
-\mathcal{L}_{int}^{NC} & = & \bar{f}(p_{1})\gamma^{\mu}\left(g_{z'L}^{f}+g_{z'R}^{f}\right)f(p_{2})\frac{-ig_{\mu\lambda}}{(p_{2}-p_{1})^{2}-M_{z'}^{2}}\bar{\nu}_{\ell L}(p_{3})\gamma^{\lambda}g_{\nu z'}\nu_{\ell L}(p_{4})\nonumber \\
 & + & \bar{f}(p_{1})\gamma^{\mu}\left(g_{zL}^{f}+g_{zR}^{f}\right)f(p_{2})\frac{-ig_{\mu\lambda}}{(p_{2}-p_{1})^{2}-M_{z}^{2}}\bar{\nu}_{\ell L}(p_{3})\gamma^{\lambda}g_{\nu z}\nu_{\ell L}(p_{4})\nonumber \\
 & + & \bar{f}(p_{1})\gamma^{\mu}\left(g_{k'L}^{f}+g_{k'R}^{f}\right)f(p_{2})\frac{-ig_{\mu\lambda}}{(p_{2}-p_{1})^{2}-M_{k'}^{2}}\bar{\nu}_{\ell L}(p_{3})\gamma^{\lambda}g_{\nu k'}\nu_{\ell L}(p_{2}).\label{eq:3.24}\end{eqnarray}
For low energies, we have that $M_{k'}^{2},\, M_{z}^{2},\, M_{z'}^{2}\gg(p_{2}-p_{1})^{2}$
with $p_{3}=p_{4}=p$ and eq. (\ref{eq:3.24}) can be written as:\begin{eqnarray}
-\mathcal{L}_{eff}^{NC} & \approx & \frac{G_{\nu z'}}{M_{z'}^{2}}\left\langle \bar{f}(p_{1})\gamma^{\mu}\left(g_{z'L}^{f}+g_{z'R}^{f}\right)f(p_{2})\right\rangle \bar{\nu}_{\ell L}\gamma_{\mu}\nu_{\ell L}\nonumber \\
 & + & \frac{G_{\nu z}}{M_{z}^{2}}\left\langle \bar{f}(p_{1})\gamma^{\mu}\left(g_{zL}^{f}+g_{zR}^{f}\right)f(p_{2})\right\rangle \bar{\nu}_{\ell L}\gamma_{\mu}\nu_{\ell L}\nonumber \\
 & + & \frac{G_{\nu k'}}{M_{k'}^{2}}\left\langle \bar{f}(p_{1})\gamma^{\mu}\left(g_{k'L}^{f}+g_{k'R}^{f}\right)f(p_{2})\right\rangle \bar{\nu}_{\ell L}\gamma_{\mu}\nu_{\ell L}\label{eq:3.25}.\end{eqnarray}
Following the same procedure of section \ref{sub:Charged-Currents}
we obtain:\begin{eqnarray}
\mathcal{L}_{eff}^{NC} & \approx & -\sum_{P=L,R}\left(g_{z'P}^{f}\frac{G_{\nu z'}}{M_{z'}^{2}}+g_{zP}^{f}\frac{G_{\nu z}}{M_{z}^{2}}+g_{k'P}^{f}\frac{G_{\nu k'}}{M_{k'}^{2}}\right)\frac{1}{2}n_{f}\bar{\nu}_{\ell L}\gamma_{0}\nu_{\ell L}.\label{eq:3.26}\end{eqnarray}

\subsubsection{Leptons sector}

From eqs. (\ref{eq:6}) and (\ref{eq:29}), we obtain that for the
known neutral leptons:\begin{eqnarray}
\frac{g_{x}}{3}\bar{\nu}_{\ell L}\gamma^{\mu}\nu_{\ell L}B_{\mu} & = & \bar{\nu}_{\ell L}\gamma^{\mu}\nu_{\ell L}\left[-\frac{g}{3}S_{W}A_{\mu}+\left(\frac{g}{3}T_{W}^{2}C_{W}+\frac{g_{x}}{3}\beta_{1}\right)Z_{\mu}^{0}\right.\nonumber \\
 &  & \left.-\frac{g_{x}}{3}\left(\frac{1}{\sqrt{3}}T_{W}-\beta_{2}\right)K_{{R}\mu}^{'0}\right],\label{eq:3.7}\\
\frac{g}{2}\bar{\nu}_{\ell L}\gamma^{\mu}\nu_{\ell L}W_{3}^{\mu} & = & \bar{\nu}_{\ell L}\gamma^{\mu}\nu_{\ell L}\left[\frac{g}{2}S_{W}A_{\mu}-\frac{g\vartheta_{1}}{2V}Z_{\mu}^{'0}+\frac{g\left(C_{W}+\beta_{3}\right)}{2}Z_{\mu}^{0}+\frac{g\beta_{4}}{2}K_{{R}\mu}^{'0}\right]\,\,\,\,\,\,\,\,\,\,\,\, ,\label{eq:3.8}\\
\frac{-g}{2\sqrt{3}}\bar{\nu}_{\ell L}\gamma^{\mu}\nu_{\ell L}W_{8}^{\mu} & = & \bar{\nu}_{\ell L}\gamma^{\mu}\nu_{\ell L}\left[\frac{-g}{6}S_{W}A_{\mu}-\frac{g\vartheta_{1}}{2V}Z_{\mu}^{'0}+\left(\frac{g}{6}\frac{S_{W}^{2}}{C_{W}}-\frac{g\beta_{5}}{2\sqrt{3}}\right)Z_{\mu}^{0}\right.\nonumber \\
 &  & \left.+\frac{g}{2\sqrt{3}}\left(\frac{1}{t}T_{W}-\beta_{6}\right)K_{{R}\mu}^{'0}\right]\label{eq:3.9}.\end{eqnarray}
By eqs. (\ref{eq:3.7}), (\ref{eq:3.8}) and (\ref{eq:3.9}) we obtain
that vertex interactions with neutrinos can be written as:\begin{eqnarray}
\bar{\nu}_{\ell L}\gamma^{\mu}\nu_{\ell L}A_{\mu} & \propto & 0~,\label{eq:3.10}\\
\bar{\nu}_{\ell L}\gamma^{\mu}\nu_{\ell L}Z_{\mu}^{'0} & \propto & -\frac{g\vartheta_{1}}{V}\equiv G_{\nu Z'}\label{eq:3.11},\\
\bar{\nu}_{\ell L}\gamma^{\mu}\nu_{\ell L}Z_{\mu}^{0} & \propto & \frac{1}{2}gC_{W}^{-1}+\eta_{1}\equiv G_{\nu Z}\label{eq:3.12},\\
\bar{\nu}_{\ell L}\gamma^{\mu}\nu_{\ell L}K_{{R}\mu}^{'0} & \propto & \left(\frac{3g-2g_{x}t}{6\,\sqrt{3}t}\right)T_{W}+\eta_{2}\equiv G_{\nu K'},\label{eq:3.13}\end{eqnarray}
where
\begin{eqnarray*}
\eta_{1} & =&\frac{-4gtC_{W}^{2}\vartheta_{1}^{2}+g_{x}\left(1-2S_{W}^{2}\right)\vartheta_{2}^{2}}{8tV^{2}C_{W}^{5}},\\
\eta_{2} & =&\frac{gt\left(1-4C_{W}^{2}\right)\vartheta_{1}^{2}}{2\sqrt{3}V^{2}C_{W}S_{W}}-\frac{\left(-gt^{3}+2gt^{3}C_{W}^{2}+8gt^{3}C_{W}^{4}+6g_{x}S_{W}^{4}\right)\vartheta_{2}^{2}}{24\sqrt{3}t^{2}V^{2}C_{W}^{5}S_{W}}.
\end{eqnarray*}
We note from eq. (\ref{eq:3.10}) that neutrinos does not interact
electrically, as expected. For charged leptons, from eqs. (\ref{eq:6})
and (\ref{eq:29}) we obtain:\begin{eqnarray}
\frac{g_{x}}{3}\bar{\ell_{L}}\gamma^{\mu}\ell_{L}B_{\mu} & = & \bar{\ell_{L}}\gamma^{\mu}\ell_{L}\left[\frac{-g}{3}S_{W}A_{\mu}+\left(\frac{g}{3}T_{W}^{2}C_{W}+\frac{g_{x}}{3}\beta_{1}\right)Z_{\mu}^{0}\right.\nonumber \\
 &  & \left.-\frac{g_{x}}{3}\left(\frac{1}{\sqrt{3}}T_{W}-\beta_{2}\right)K_{{R}\mu}^{'0}\right]\label{eq:3.14},\\
-\frac{g}{2}\bar{\ell}_{L}\gamma^{\mu}\ell_{L}W_{3}^{\mu} & = & \bar{\ell}_{L}\gamma^{\mu}\ell_{L}\left[\frac{-g}{2}S_{W}A_{\mu}+\frac{g\vartheta_{1}}{2V}Z_{\mu}^{'0}-\frac{g\left(C_{W}+\beta_{3}\right)}{2}Z_{\mu}^{0}-\frac{g\beta_{4}}{2}K_{{R}\mu}^{'0}\right],\,\,\,\,\,\,\,\,\,\,\label{eq:3.15}\\
\frac{-g}{2\sqrt{3}}\bar{\ell}_{L}\gamma^{\mu}\ell_{L}W_{8}^{\mu} & = & \bar{\ell}_{L}\gamma^{\mu}\ell_{L}\left[\frac{-g}{6}S_{W}A_{\mu}-\frac{g\vartheta_{1}}{2V}Z_{\mu}^{'0}+\left(\frac{g}{6}\frac{S_{W}^{2}}{C_{W}}-\frac{g\beta_{5}}{2\sqrt{3}}\right)Z_{\mu}^{0}\right.\nonumber \\
 &  & \left.+\frac{g}{2\sqrt{3}}\left(\frac{1}{t}T_{W}-\beta_{6}\right)K_{{R}\mu}^{'0}\right],\,\,\,\,\,\,\,\,\,\label{eq:3.16}\\
g_{x}\bar{\ell_{R}}\gamma^{\mu}\ell_{R}B_{\mu} & = & \bar{\ell_{R}}\gamma^{\mu}\ell_{R}\left[-gS_{W}A_{\mu}+\left(gT_{W}^{2}C_{W}+g_{x}\beta_{1}\right)Z_{\mu}^{0}\right.\nonumber \\
 &  & \left.-g_{x}\left(\frac{1}{\sqrt{3}}T_{W}-\beta_{2}\right)K_{{R}\mu}^{'0}\right],\label{eq:3.16-1}\end{eqnarray}
and therefore:
\begin{eqnarray}
\bar{\ell}\gamma^{\mu}\ell A_{\mu} & \propto & -gS_{W}\label{eq:3.17},\\
\overline{\ell_{L}}\gamma^{\mu}\ell_{L}Z_{\mu}^{'0} & \propto & 0\equiv g_{z'L}^{\ell}=g_{z'R}^{\ell}\label{eq:3.19},\\
\overline{\ell_{L}}\gamma^{\mu}\ell_{L}Z_{\mu}^{0} & \propto & \frac{1}{2}g\left(-1+T_{W}^{2}\right)C_{W}+\eta_{3}\equiv g_{zL}^{\ell}\label{eq:3.20},\\
\overline{\ell_{R}}\gamma^{\mu}\ell_{R}Z_{\mu}^{0} & \propto & gT_{W}^{2}C_{W}+\eta_{5}\equiv g_{zR}^{\ell}\label{eq:3.20-1},\\
\overline{\ell_{L}}\gamma^{\mu}\ell_{L}K_{{R}\mu}^{'0} & \propto & \frac{1}{6\sqrt{3}t}\left(3g-2tg_{x}\right)T_{W}+\eta_{4}\equiv g_{k'L}^{\ell}\label{eq:3.22},\\
\overline{\ell_{R}}\gamma^{\mu}\ell_{R}K_{{R}\mu}^{'0} & \propto & -\frac{g_{x}}{\sqrt{3}}T_{W}+\eta_{6}\equiv g_{k'R}^{\ell}\label{eq:3.22-1},\end{eqnarray}
where 
\begin{eqnarray*}
\eta_{3} & =&\frac{\left(-1+2C_{W}^{2}\right)g_{x}\vartheta_{2}^{2}}{8tV^{2}C_{W}^{5}},\\
\eta_{4} & =&\frac{\left(gt^{3}\left(1+2C_{W}^{2}\right)^{2}-12gt^{3}S_{W}^{2}C_{W}^{2}-6g_{x}S_{W}^{4}\right)}{24\sqrt{3}t^{2}V^{2}C_{W}^{5}S_{W}},\\
\eta_{5} & =&-\frac{g_{x}\vartheta_{2}^{2}}{4tV^{2}C_{W}^{3}}T_{W}^{2},\\
\eta_{6} & =&-\frac{\sqrt{3}g_{x}\vartheta_{2}^{2}}{4t^{2}V^{2}C_{W}^{2}}T_{W}^{3},
\end{eqnarray*}	
and, again, $t=g_{x}/g$. 
We note that by eq. (\ref{eq:3.17}) we can make the association
$gS_{W}=|e|$. Then for $f=e$, eqs. (\ref{eq:3.11})-(\ref{eq:3.13})
and (\ref{eq:3.19})-(\ref{eq:3.22-1}) lead to:

\begin{eqnarray}
\mathcal{L}_{eff-e}^{NC} & \approx & -\sum_{P=L,R}\frac{1}{2}\left(g_{z'P}^{e}\frac{G_{\nu z'}}{M_{z'}^{2}}+g_{zP}^{e}\frac{G_{\nu z}}{M_{z}^{2}}+g_{k'P}^{e}\frac{G_{\nu k'}}{M_{k'}^{2}}\right)n_{e}\bar{\nu}_{\ell L}\gamma_{0}\nu_{\ell L}\nonumber \\
 & \approx & -\left\{ \left[\frac{T_{W}^{4}}{144t^{2}g_{x}^{2}V^{2}}\left(3g-2tg_{x}\right)^{2}+\frac{T_{W}^{2}}{8V^{2}}\left(1-T_{W}^{2}\right)\right.\right.\nonumber \\
 & + & \left.\frac{1}{2}\left(\frac{1}{2\vartheta_{2}^{2}}-\frac{\vartheta_{1}^{2}}{2V^{2}\vartheta_{2}^{2}}\right)\left(1-2C_{W}^{2}\right)\right]_{L}\nonumber \\
 & + & \left.\left[\frac{T_{W}^{4}\left(2tg_{x}-3g\right)}{24tg_{x}V^{2}}-\frac{T_{W}^{4}}{4V^{2}}+\left(\frac{1}{2\vartheta_{2}^{2}}-\frac{\vartheta_{1}^{2}}{2V^{2}\vartheta_{2}^{2}}\right)S_{W}^{2}\right]_{R}\right\} n_{e}\bar{\nu}_{\ell L}\gamma_{0}\nu_{\ell L}.\,\,\,\,\,\,\,\,\,\,\,\label{eq:3.26-3}\end{eqnarray}
Since intermediate neutral bosons in eq. (\ref{eq:3.24}) does not
distinguish between different lepton flavors, the interaction through
NC with electron is described by the following effective potential.\begin{eqnarray}
V_{NC}^{e} & = & V_{NC}^{\mu}=V_{NC}^{\tau}=V_{NC}^{\ell},\label{eq:3.26-1}\\
 & = & V_{NC}^{\ell L}+V_{NC}^{\ell R},\label{eq:3.26-0}\end{eqnarray}
where\begin{eqnarray}
V_{NC}^{\ell L} & = & \left[\frac{T_{W}^{4}}{144t^{2}g_{x}^{2}V^{2}}\left(3g-2tg_{x}\right)^{2}+\frac{T_{W}^{2}}{8V^{2}}\left(1-T_{W}^{2}\right)\right.\nonumber \\
 &  & +\left.\frac{1}{2}\left(\frac{1}{2\vartheta_{2}^{2}}-\frac{\vartheta_{1}^{2}}{V^{2}\vartheta_{2}^{2}}\right)\left(1-2C_{W}^{2}\right)\right]n_{e},\label{eq:326}\\
V_{NC}^{\ell R} & = & \left[\frac{T_{W}^{4}\left(2tg_{x}-3g\right)}{24tg_{x}V^{2}}-\frac{T_{W}^{4}}{4V^{2}}+\left(\frac{1}{2\vartheta_{2}^{2}}-\frac{\vartheta_{1}^{2}}{2V^{2}\vartheta_{2}^{2}}\right)S_{W}^{2}\right]n_{e},\label{eq:327}\end{eqnarray}
and index $\ell$ refers to neutrino flavor. We note that
the potential through CC comes from interactions of electron neutrinos
with left-handed electrons, while the effective potential through
NC comes from left and right-handed electrons.

Considering both NC and CC, we can write the effective potential felt
by neutrinos as $V^{\ell}=V^{\ell L}+V^{\ell R}$ where\begin{eqnarray}
V^{\ell L} & = & \left(\frac{1}{2\vartheta_{2}^{2}}-\frac{\vartheta_{1}^{2}}{2V^{2}\vartheta_{2}^{2}}\right)\delta_{e\ell}n_{e}+V_{NC}^{\ell L},\label{eq:3.27-1}\\
V^{\ell R} & = & V_{NC}^{\ell R}\label{eq:3.27-2}.\end{eqnarray}
Comparing with SM expression for such potential:\begin{eqnarray}
\mathit{\mathsf{{V}}}_{NC}^{\ell} & = & -\sqrt{2}G_{F}\left(\frac{1}{2}-2S_{W}^{2}\right)n_{e},\,\,\mathit{\mathsf{{V}}}_{CC}^{e}=\sqrt{2}G_{F}n_{e},\label{eq:3.28}\end{eqnarray}
we can find that\begin{eqnarray}
V^{\ell L} & = & \mathit{\mathsf{{V}}}^{\ell L}+\left[\frac{T_{W}^{4}}{144t^{2}g_{x}^{2}V^{2}}\left(3g-2tg_{x}\right)^{2}+\frac{T_{W}^{2}}{8V^{2}}\left(1-T_{W}^{2}\right)\right]n_{e},\label{eq:3.29-4}\\
V^{\ell R} & = & \mathit{\mathsf{{V}}}_{NC}^{\ell R}+\left[\frac{T_{W}^{4}\left(2tg_{x}-3g\right)}{24tg_{x}V^{2}}-\frac{T_{W}^{4}}{4V^{2}}\right]n_{e},\label{eq:3.29-5}\end{eqnarray}
then, the new terms beyond SM $\left[\,\right]$
can be associated with the parameters $\varepsilon'$s
in NSI~\cite{int-31}. So, in the approximation $\left(\frac{\vartheta_{i}}{V}\right)^{n}\approx0$,
for $n>2$ we obtain:\begin{eqnarray}
\varepsilon_{\ell\ell}^{eL} & \approx & \frac{\left(1-2S_{W}^{2}\right)\vartheta_{2}^{2}}{8V^{2}C_{W}^{4}}\label{eq:eL},\\
\varepsilon_{\ell\ell}^{eR} & \approx & -\frac{S_{W}^{2}\left(1+2S_{W}^{2}\right)\vartheta_{2}^{2}}{4V^{2}C_{W}^{4}}\label{eq:eR}.\end{eqnarray}
We note that on limit $V\rightarrow\infty$ we recover SM. The NSI
are a sub-leading interaction, as expected. 
By eq.~(\ref{eq:eL}) and eq.~(\ref{eq:eR}) we obtain: 
$\varepsilon_{\ell\ell}^{eR}\approx-2S_{W}^{2}\varepsilon_{\ell\ell}^{eL}-\frac{\vartheta_{2}^{2}}{V^{2}}T_{W}^{4}$.

\subsubsection{Quarks sector}

For the quarks of first family, Lagrangian density in eq.~(\ref{eq:16})
describes the interactions with gauge bosons $W_{3\mu},\, W_{8\mu}$and
$B_{\mu}$, then by eqs.~(\ref{eq:29}) and (\ref{eq:31}) we obtain
the following interactions for quarks up:
\begin{eqnarray}
-\frac{g_{x}}{3}\overline{u_{L}}\gamma^{\mu}u_{L}B_{\mu} & = & \overline{u_{L}}\gamma^{\mu}u_{L}\left[\frac{g}{3}S_{W}A_{\mu}-\frac{g_{x}}{3}\left(\frac{1}{t}T_{W}^{2}C_{W}+\beta_{1}\right)Z_{\mu}^{0}\right.\nonumber \\
 &  & +\left.\frac{g_{x}}{3}\left(\frac{1}{\sqrt{3}}T_{W}-\beta_{2}\right)K_{{R}\mu}^{'0}\right],\label{eq:3.30}\\
\frac{g}{2}\overline{u_{L}}\gamma^{\mu}u_{L}W_{3}^{\mu} & = & \overline{u_{L}}\gamma^{\mu}u_{L}\left[\frac{g}{2}S_{W}A_{\mu}-\frac{g\vartheta_{1}}{2V}Z_{\mu}^{'0}+\frac{g\left(C_{W}+\beta_{3}\right)}{2}Z_{\mu}^{0}+\frac{g\beta_{4}}{2}K_{{R}\mu}^{'0}\right],\,\,\,\,\,\,\label{eq:3.31}\\
\frac{-g}{2\sqrt{3}}\overline{u_{L}}\gamma^{\mu}u_{L}W_{8}^{\mu} & = & \overline{u_{L}}\gamma^{\mu}u_{L}\left[\frac{-g}{6}S_{W}A_{\mu}-\frac{g\vartheta_{1}}{2V}Z_{\mu}^{'0}\right.\nonumber \\
 &  & +\left.\frac{g}{2\sqrt{3}}\left(\frac{1}{\sqrt{3}}T_{W}S_{W}-\beta_{5}\right)Z_{\mu}^{0}+\frac{g}{2\sqrt{3}}\left(\frac{1}{t}T_{W}-\beta_{6}\right)K_{{R}\mu}^{'0}\right],\label{eq:3.32}\\
-\frac{2g_{x}}{3}\overline{u_{R}}\gamma^{\mu}u_{R}B_{\mu} & = & \overline{u_{R}}\gamma^{\mu}u_{R}\left[\frac{2g}{3}S_{W}A_{\mu}-\frac{2g_{x}}{3}\left(\frac{1}{t}T_{W}^{2}C_{W}+\beta_{1}\right)Z_{\mu}^{0}\right.\nonumber \\
 &  & +\left.\frac{2g_{x}}{3}\left(\frac{1}{t}T_{W}-\beta_{6}\right)K_{{R}\mu}^{'0}\right].\label{eq:3.32-1}\end{eqnarray}
The couplings quark-quark-boson for the first family are given by:
\begin{eqnarray}
\overline{u_{L}}\gamma^{\mu}u_{L}A_{\mu} & \propto & \frac{2}{3}gS_{W},\label{eq:3.36}\\
\overline{u_{R}}\gamma^{\mu}u_{R}A_{\mu} & \propto & \frac{2}{3}gS_{W},\label{eq:3.36-1}\\
\overline{u_{L}}\gamma^{\mu}u_{L}Z_{\mu}^{'0} & \propto & -\frac{g\vartheta_{1}}{V}\equiv g_{z'L}^{u}\label{eq:3.37},\\
\overline{u_{R}}\gamma^{\mu}u_{R}Z_{\mu}^{'0} & \propto & 0\equiv g_{z'R}^{u}\label{eq:3.37-1},\\
\overline{u_{L}}\gamma^{\mu}u_{L}Z_{\mu}^{0} & \propto & \frac{1}{6}g\left(3-T_{W}^{2}\right)C_{W}+\zeta_{1}\equiv g_{zL}^{u}\label{eq:3.38},\\
\overline{u_{R}}\gamma^{\mu}u_{R}Z_{\mu}^{0} & \propto & -\frac{2}{3}gT_{W}^{2}C_{W}+\zeta_{3}\equiv g_{zR}^{u}\label{eq:3.38-1},\\
\overline{u_{L}}\gamma^{\mu}u_{L}K_{{R}\mu}^{'0} & \propto & \frac{1}{6\sqrt{3}t}\left(3g+2tg_{x}\right)T_{W}\equiv g_{k'L}^{u},\label{eq:3.39}\\
\overline{u_{R}}\gamma^{\mu}u_{R}K_{{R}\mu}^{'0} & \propto & \frac{2}{3\sqrt{3}}g_{x}T_{W}+\zeta_{4}\equiv g_{k'R}^{u}\label{eq:3.39-1},
\end{eqnarray}
where
\begin{eqnarray}
\zeta_{1} & =&\frac{g_{x}\left(-12C_{W}^{4}\vartheta_{1}^{2}+\left(1+2C_{W}^{2}\right)\vartheta_{2}^{2}\right)}{24tV^{2}C_{W}^{5}},\nonumber \\
\zeta_{2} & =&\frac{12gt^{3}C_{W}^{4}\left(1-4C_{W}^{2}\right)\vartheta_{1}^{2}+\left(gt^{3}\left(1-2C_{W}^{2}-8C_{W}^{4}\right)+6g_{x}S_{W}^{4}\right)\vartheta_{2}^{2}}{24\sqrt{3}t^{2}V^{2}C_{W}^{5}S_{W}},\nonumber \\
\zeta_{3} & =&\frac{g}{6}\frac{S_{W}^{2}\vartheta_{2}^{2}}{C_{W}^{5}V^{2}},\nonumber \\
\zeta_{4} & =&\frac{g_{x}S_{W}^{3}\vartheta_{2}^{2}}{2\sqrt{3}t^{2}V^2C_{W}^{5}}.\label{eq:-9}
\end{eqnarray}
We note that eqs. (\ref{eq:3.36}) and (\ref{eq:3.36-1}) reflect
the fact that quarks interact electrically through photons with coupling
constant $Q_{f}\sin\theta_{W}$, as in SM. The effective lagrangian
at low energies for neutrino interaction with quarks up through neutral
currents are given by eq. \ref{eq:3.26} with $f=u$:
\begin{eqnarray}
\mathcal{L}_{quark,\, u}^{NC} & \approx & -\frac{1}{2}\sum_{P=L,R}\left(g_{z'P}^{u}\frac{G_{\nu z'}}{M_{z'}^{2}}+g_{zP}^{u}\frac{G_{\nu z}}{M_{z}^{2}}+g_{k'P}^{u}\frac{G_{\nu k'}}{M_{k'}^{2}}\right)n_{u}\bar{\nu}_{\ell L}\gamma_{0}\nu_{\ell L}\nonumber \\
 & \approx & -\left\{ \left[\frac{1}{24V^{2}}\left(3+T_{W}^{4}\right)+\frac{T_{W}^{4}}{144t^{4}V^{2}}\left(9-4t^{4}\right)\right.\right.\nonumber \\
 & + & \left.\left(\frac{1}{2\vartheta_{2}^{2}}-\frac{\vartheta_{1}^{2}}{2V^{2}\vartheta_{2}^{2}}\right)\left(\frac{1}{2}-\frac{2}{3}S_{W}^{2}\right)-\frac{\vartheta_{1}^{2}}{4V^{2}\vartheta_{2}^{2}}\right]_{L}\nonumber \\
 & + & \left.\left[\frac{T_{W}^{4}}{6V^{2}}+\frac{T_{W}^{4}\left(3g-2tg_{x}\right)}{36tg_{x}V^{2}}-\frac{2}{3}\left(\frac{1}{2\vartheta_{2}^{2}}-\frac{\vartheta_{1}^{2}}{2V^{2}\vartheta_{2}^{2}}\right)S_{W}^{2}\right]_{R}\right\} n_{u}\bar{\nu}_{\ell L}\gamma_{0}\nu_{\ell L},\,\,\,\,\,\,\,\,\,\,\,\,\,\,\,\label{eq:3.44-1}\end{eqnarray}
where $n_{u}$ is the up-quarks average density. 

SM predictions, using result of eq.(~\ref{eq:3.6}), can be
written as:\begin{eqnarray}
\mathit{\mathsf{{V}}}_{NC}^{u} & = & \mathit{\mathsf{{V}}}_{NC}^{uL}+\mathit{\mathsf{{V}}}_{NC}^{uR}=\left(\frac{1}{2\vartheta_{2}^{2}}-\frac{\vartheta_{1}^{2}}{2V^{2}\vartheta_{2}^{2}}\right)\left(\frac{1}{2}-\frac{4}{3}S_{W}^{2}\right)n_{u},\end{eqnarray}
where\begin{eqnarray}
\mathit{\mathsf{{V}}}_{NC}^{uL} & = & \left(\frac{1}{2\vartheta_{2}^{2}}-\frac{\vartheta_{1}^{2}}{2V^{2}\vartheta_{2}^{2}}\right)\left(\frac{1}{2}-\frac{2}{3}S_{W}^{2}\right)n_{u}\label{eq:3.45},\\
\mathit{\mathsf{{V}}}_{NC}^{uR} & = & -\frac{2}{3}\left(\frac{1}{2\vartheta_{2}^{2}}-\frac{\vartheta_{1}^{2}}{2V^{2}\vartheta_{2}^{2}}\right)S_{W}^{2}n_{u}.\label{eq:3.45-0}\end{eqnarray}
By comparison, we obtain:\begin{eqnarray}
V_{NC}^{uL} & \approx & \mathit{\mathsf{{V}}}_{NC}^{uL}+\left[\frac{1}{24V^{2}}\left(3+T_{W}^{4}\right)+\frac{T_{W}^{4}}{144t^{4}V^{2}}\left(9-4t^{4}\right)-\frac{\vartheta_{1}^{2}}{4V^{2}\vartheta_{2}^{2}}\right]n_{u}\label{eq:3.45-1},\\
V_{NC}^{uR} & \approx & \mathit{\mathsf{{V}}}_{NC}^{uR}+\left[\frac{T_{W}^{4}}{6V^{2}}+\frac{T_{W}^{4}\left(3g-2tg_{x}\right)}{36tg_{x}V^{2}}\right]n_{u}.\label{eq:3.45-2}\end{eqnarray}
Then we can say that $\varepsilon_{\ell\ell}^{u}=\varepsilon_{\ell\ell}^{uL}+\varepsilon_{\ell\ell}^{uR}$
where\begin{eqnarray}
\varepsilon_{\ell\ell}^{uL} & \approx & -\frac{\vartheta_{1}^{2}}{2V^{2}}+\frac{\vartheta_{2}^{2}}{24V^{2}C_{W}^{4}}\left(9-8S_{W}^{2}\right),\label{eq:uL}\\
\varepsilon_{\ell\ell}^{uR} & \approx & \frac{\vartheta_{2}^{2}}{6V^{2}}\frac{S_{W}^{2}}{C_{W}^{4}}\label{eq:uR},\end{eqnarray}
again, we obtain universal NSI, as for the electrons. We note that
$\varepsilon_{\ell\ell}^{uL}=-\frac{\vartheta_{1}^{2}}{2V^{2}}+\frac{3\vartheta_{2}^{2}}{8V^{2}C_{W}^{4}}-2\varepsilon_{\ell\ell}^{uR}$
and in the limit $V\rightarrow\infty$ we recover SM.

For quarks down by eq. (\ref{eq:16}) and (\ref{eq:29}) we obtain
that:\begin{eqnarray}
-\frac{g_{x}}{3}\overline{d_{L}}\gamma^{\mu}d_{L}B_{\mu} & = & \overline{d_{L}}\gamma^{\mu}d_{L}\left[\frac{g}{3}S_{W}A_{\mu}-\frac{g_{x}}{3}\left(\frac{1}{t}T_{W}^{2}C_{W}+\beta_{1}\right)Z_{\mu}^{0}\right.\nonumber \\
 &  & +\left.\frac{g_{x}}{3}\left(\frac{1}{\sqrt{3}}T_{W}-\beta_{2}\right)K_{{R}\mu}^{'0}\right],\label{eq:3.33}\\
-\frac{g}{2}\overline{d_{L}}\gamma^{\mu}d_{L}W_{3}^{\mu} & = & \overline{d_{L}}\gamma^{\mu}d_{L}\left[-\frac{gS_{W}}{2}A_{\mu}+\frac{g\vartheta_{1}}{2V}Z_{\mu}^{'0}-\frac{g\left(C_{W}+\beta_{3}\right)}{2}Z_{\mu}^{0}-\frac{g\beta_{4}}{2}K_{{R}\mu}^{'0}\right],\,\,\,\,\,\,\,\,\,\,\,\,\,\,\,\,\label{eq:3.34}\\
\frac{-g}{2\sqrt{3}}\overline{d_{L}}\gamma^{\mu}d_{L}W_{8}^{\mu} & = & \overline{d_{L}}\gamma^{\mu}d_{L}\left[\frac{-gS_{W}}{6}A_{\mu}+\frac{g}{2\sqrt{3}}\left(\frac{1}{\sqrt{3}}T_{W}S_{W}-\beta_{5}\right)Z_{\mu}^{0}\right.\nonumber \\
 &  & +\left.-\frac{g\vartheta_{1}}{2V}Z_{\mu}^{'0}+\frac{g}{2\sqrt{3}}\left(\frac{1}{t}T_{W}-\beta_{6}\right)K_{{R}\mu}^{'0}\right],\label{eq:3.35}\\
\frac{g_{x}}{3}\overline{d_{R}}\gamma^{\mu}d_{R}B_{\mu} & = & \overline{d_{L}}\gamma^{\mu}d_{L}\left[-\frac{gS_{W}}{3}A_{\mu}+\frac{g_{x}}{3}\left(\frac{1}{t}T_{W}^{2}C_{W}+\beta_{1}\right)Z_{\mu}^{0}\right.\nonumber \\
 &  & +\left.\frac{g_{x}}{3}\left(-\frac{1}{\sqrt{3}}T_{W}+\beta_{2}\right)\right]K_{{R}\mu}^{'0}.\label{eq:3.35-1}\end{eqnarray}
Then the couplings quark-quark-boson for the first family are given
by:\begin{eqnarray}
\overline{d_{L}}\gamma^{\mu}d_{L}A_{\mu} & \propto & -\frac{1}{3}gS_{W},\label{eq:3.40}\\
\overline{d_{R}}\gamma^{\mu}d_{R}A_{\mu} & \propto & -\frac{1}{3}gS_{W},\label{eq:3.40-1}\\
\overline{d_{L}}\gamma^{\mu}d_{L}Z_{\mu}^{'0} & \propto & 0\equiv g_{z'L}^{d},\label{eq:3.41}\\
\overline{d_{R}}\gamma^{\mu}d_{R}Z_{\mu}^{'0} & \propto & 0\equiv g_{z'R}^{d},\label{eq:3.41-1}\\
\overline{d_{L}}\gamma^{\mu}d_{L}Z_{\mu}^{0} & \propto & -\frac{1}{6}g\left(3+T_{W}^{2}\right)C_{W}+\zeta_{5}\equiv g_{zL}^{d},\label{eq:3.42}\\
\overline{d_{R}}\gamma^{\mu}d_{R}Z_{\mu}^{0} & \propto & \frac{g}{3}T_{W}^{2}C_{W}+\zeta_{7}\equiv g_{zR}^{d}\label{eq:3.42-1},\\
\overline{d_{L}}\gamma^{\mu}d_{L}K_{{R}\mu}^{'0} & \propto & \frac{1}{6\sqrt{3}t}\left(3g+2tg_{x}\right)T_{W}+\zeta_{6}\equiv g_{k'L}^{d}\label{eq:3.43},\\
\overline{u_{R}}\gamma^{\mu}u_{R}K_{{R}\mu}^{'0} & \propto & -\frac{1}{3\sqrt{3}}g_{x}T_{W}+\zeta_{8}\equiv g_{k'R}^{d}\label{eq:3.43-1},\end{eqnarray}
where \begin{eqnarray}
\zeta_{5} & =&\frac{g\vartheta_{2}^{2}}{24V^{2}C_{W}^{5}}\left(3-2S_{W}^{2}\right),\nonumber \\
\zeta_{6} & =&\frac{\left(-1+3C_{W}^{2}+6C_{W}^{4}-8C_{W}^{6}\right)}{24\sqrt{3}V^{2}C_{W}^{5}S_{W}^{3}},\nonumber \\
\zeta_{7} & =&-\frac{gS_{W}^{2}\vartheta_{2}^{2}}{12V^{2}C_{W}^{5}}\nonumber ,\\
\zeta_{8} & =&-\frac{g_{x}S_{W}^{3}\vartheta_{2}^{2}}{4\sqrt{3}t^{2}V^{2}C_{W}^{5}},\label{eq:-10}\end{eqnarray}
then by eq. (\ref{eq:3.26}) for $f=d$ we obtain the following effective
lagrangian for NC:\begin{eqnarray}
\mathcal{L}_{quark,\, d}^{NC} & \approx & -\left(g_{z'V}^{d}\frac{G_{\nu z'}}{M_{z'}^{2}}+g_{zV}^{d}\frac{G_{\nu z}}{M_{z}^{2}}+g_{k'V}^{d}\frac{G_{\nu k'}}{M_{k'}^{2}}\right)n_{d}\bar{\nu}_{\ell L}\gamma_{0}\nu_{\ell L}\nonumber \\
 & \approx & -\left\{ \left[\frac{\left(3S_{W}^{2}-2S_{W}^{4}\right)}{24V^{2}C_{W}^{4}}+\frac{\left(9-4t^{4}\right)}{144t^{4}V^{2}}T_{W}^{4}\right.\right.\nonumber \\
 &  & +\left.\left(\frac{1}{2\vartheta_{2}^{2}}-\frac{\vartheta_{1}^{2}}{2V^{2}\vartheta_{2}^{2}}\right)\left(-\frac{1}{2}+\frac{1}{3}S_{W}^{2}\right)\right]_{L}\\
 &  & \left.+\left[-\frac{S_{W}^{2}}{24V^{2}C_{W}^{4}}+\frac{1}{3}\left(\frac{1}{2\vartheta_{2}^{2}}-\frac{\vartheta_{1}^{2}}{2V^{2}\vartheta_{2}^{2}}\right)S_{W}^{2}\right]_{R}\right\} n_{d}\bar{\nu}_{\ell L}\gamma_{0}\nu_{\ell L}\label{eq:3.46},\end{eqnarray}
and the effective potential felt by neutrinos when crossing a medium
composed by a density $n_{d}$ ~of~ $down$ quarks is $V_{NC}^{d}=V_{NC}^{dL}+V_{NC}^{dR}$
where\begin{eqnarray}
V_{NC}^{dL} & \approx & \left[\frac{\left(3S_{W}^{2}-2S_{W}^{4}\right)}{24V^{2}C_{W}^{4}}+\frac{\left(9-4t^{4}\right)}{144t^{4}V^{2}}T_{W}^{4}+\left(\frac{1}{2\vartheta_{2}^{2}}-\frac{\vartheta_{1}^{2}}{2V^{2}\vartheta_{2}^{2}}\right)\left(-\frac{1}{2}+\frac{1}{3}S_{W}^{2}\right)\right]n_{d},\,\,\,\,\,\,\,\,\,\,\,\,\label{eq:3.47-1}\\
V_{NC}^{dR} & \approx & \left[-\frac{S_{W}^{2}}{24V^{2}C_{W}^{4}}+\frac{1}{3}\left(\frac{1}{2\vartheta_{2}^{2}}-\frac{\vartheta_{1}^{2}}{2V^{2}\vartheta_{2}^{2}}\right)S_{W}^{2}\right]n_{d}.\label{eq:3.47-2}\end{eqnarray}
Then we can easily see that in SM the NC effective potential for neutrinos
in a d-quark medium, using result of eq.~\ref{eq:3.6}, 
will be given by:\begin{eqnarray}
\mathit{\mathsf{{V}}}_{NC}^{d} & = & \mathit{\mathsf{{V}}}_{NC}^{dL}+\mathit{\mathsf{{V}}}_{NC}^{dR}\approx-\left(\frac{1}{2\vartheta_{2}^{2}}-\frac{\vartheta_{1}^{2}}{2V^{2}\vartheta_{2}^{2}}\right)\left(\frac{1}{2}-\frac{2}{3}S_{W}^{2}\right)n_{d},\nonumber \\
\mathit{\mathsf{{V}}}_{NC}^{dL} & = & \left(\frac{1}{2\vartheta_{2}^{2}}-\frac{\vartheta_{1}^{2}}{2V^{2}\vartheta_{2}^{2}}\right)\left(-\frac{1}{2}+\frac{1}{3}S_{W}^{2}\right)n_{d},\label{eq:3.48}\\
\mathit{\mathsf{{V}}}_{NC}^{dR} & = & \frac{1}{3}\left(\frac{1}{2\vartheta_{2}^{2}}-\frac{\vartheta_{1}^{2}}{2V^{2}\vartheta_{2}^{2}}\right)S_{W}^{2}n_{d}.\label{eq:3.48-1}\end{eqnarray}
Then from eq. (\ref{eq:3.47-1})- (\ref{eq:3.48-1}), we obtain:\begin{eqnarray}
V_{CN}^{dL} & \approx & \mathit{\mathsf{{V}}}_{NC}^{dL}+\left[\frac{\left(3S_{W}^{2}-2S_{W}^{4}\right)}{24V^{2}C_{W}^{4}}+\frac{\left(9-4t^{4}\right)}{144t^{4}V^{2}}T_{W}^{4}\right]n_{d},\label{eq:3.47-3}\\
V_{NC}^{dR} & \approx & \mathit{\mathsf{{V}}}_{NC}^{dR}-\frac{S_{W}^{2}}{24V^{2}C_{W}^{4}}n_{d}\label{eq:3.47-4},\end{eqnarray}
and neglecting terms of order $\left(\frac{\vartheta_{i}}{V}\right)^{n},$
for $n>2$ we obtain that $\varepsilon_{\ell\ell}^{d}=\varepsilon_{\ell\ell}^{dL}+\varepsilon_{\ell\ell}^{dR}$
where \begin{eqnarray}
\varepsilon_{\ell\ell}^{dL} & \approx & \frac{\vartheta_{2}^{2}}{24V^{2}C_{W}^{4}}\left(3-2S_{W}^{2}\right),\label{eq:dL}\\
\varepsilon_{\ell\ell}^{dR} & \approx & -\frac{S_{W}^{2}\vartheta_{2}^{2}}{12V^{2}C_{W}^{4}}.\label{eq:dR}\end{eqnarray}
Then we obtain $\varepsilon_{\ell\ell}^{dL}\approx\frac{\vartheta_{2}^{2}}{8V^{2}C_{W}^{4}}+\varepsilon_{\ell\ell}^{dR}$.
Note that again in limit $V\rightarrow\infty$ we recover the SM.

\section{RESULTS}

In last sections we saw that in 331 model we chosed, all NSI parameters
are universal and diagonal, and will not affect oscillation experiments.
However, measurements of cross-section will be sensitive to such parameters,
through modifications on $g_{i}^{\alpha}$\cite{nsic-2}. We will
now compare our results with those obtained in cross-section measurements.
We will assume $\sin^{2}\theta_{W}=0.23149(13)$.

\begin{table}[t]
\caption{Values for NSI in 331 model and experimental limits}
\begin{center}
\begin{tabular*}{0.75\textwidth}{@{\extracolsep{\fill}}cc|c}
\hline 
\multicolumn{2}{c}{Modelo331} & Exp. 90\% C.L.~\cite{key-2} \\
\hline \hline
\multirow{3}{*}{$\varepsilon_{\ell\ell}^{eL}\approx
\frac{\left(1-2S_{W}^{2}\right)\vartheta_{2}^{2}}{8V^{2}C_{W}^{4}}$} &  
\multirow{3}{*}{$0.114\left(\frac{\vartheta_{2}^{2}}{V^{2}}\right)$}&
$-0.07<\varepsilon_{ee}^{eL}<0.11$\\
&&$-0.025<\varepsilon_{\mu\mu}^{eL}<0.03$\\
&&$-0.6<\varepsilon_{\tau\tau}^{eL}<0.4$\\
\hline
\multirow{3}{*}{$\varepsilon_{\ell\ell}^{eR}\approx
-2S_{W}^{2}\varepsilon_{\ell\ell}^{eL}-\frac{\vartheta_{2}^{2}}{V^{2}}T_{W}^{4}$}&
\multirow{3}{*}{$0.143\left(\frac{\vartheta_{2}^{2}}{V^{2}}\right)$}&
$-1<\varepsilon_{ee}^{eR}<0.5$\\
&&$-0.027<\varepsilon_{\mu\mu}^{eR}<0.03$\\
&&$-0.4<\varepsilon_{\tau\tau}^{eR}<0.6$\\
\hline
\multirow{3}{*}{$\varepsilon_{\ell\ell}^{uL}\approx-\frac{\vartheta_{1}^{2}}
{2V^{2}}+\frac{\vartheta_{2}^{2}}{24V^{2}C_{W}^{4}}\left(9-8S_{W}^{2}\right)$}&
\multirow{3}{*}{$0.50\left(\frac{\vartheta_{2}^{2}-\vartheta_{1}^{2}}{V^{2}}
\right)$}&
$-1<\varepsilon_{ee}^{uL}<0.3$\\
&&$|\varepsilon_{\mu\mu}^{uL}|<0.003$\\
&&$|\varepsilon_{\tau\tau}^{uL}|<1.4$\\
\hline
\multirow{3}{*}{$\varepsilon_{\ell\ell}^{uR}\approx\frac{\vartheta_{2}^{2}}
{6V^{2}}\frac{S_{W}^{2}}{C_{W}^{4}}$}&
\multirow{3}{*}{$0.065\left(\frac{\vartheta_{2}^{2}}{V^{2}}\right)$}&
$-0.4<\varepsilon_{ee}^{uR}<0.7$\\
&&$-0.008<\varepsilon_{\mu\mu}^{uR}<0.003$\\
&&$|\varepsilon_{\tau\tau}^{uR}|<3$\\
\hline
\multirow{3}{*}{$\varepsilon_{\ell\ell}^{dL}\approx\frac{\vartheta_{2}^{2}}
{24V^{2}C_{W}^{4}}\left(3-2S_{W}^{2}\right)$}&
\multirow{3}{*}{$0.179\left(\frac{\vartheta_{2}^{2}}{V^{2}}\right)$}&
$-0.3<\varepsilon_{ee}^{dL}<0.3$\\
&&$|\varepsilon_{\mu\mu}^{dL}|<0.003$\\
&&$|\varepsilon_{\tau\tau}^{dL}|<1.1$\\
\hline
\multirow{3}{*}{$\varepsilon_{\ell\ell}^{dR}\approx-\frac{S_{W}^{2}
\vartheta_{2}^{2}}{12V^{2}C_{W}^{4}}$}&
\multirow{3}{*}{$-0.033\left(\frac{\vartheta_{2}^{2}}{V^{2}}\right)$}&
$-0.6<\varepsilon_{ee}^{dR}<0.5$\\
&&$-0.008<\varepsilon_{\mu\mu}^{dR}<0.015$\\
&&$|\varepsilon_{\tau\tau}^{dR}|<6$\\
\hline\hline
\label{Flo:final}
\end{tabular*}
\end{center}
\end{table}

In table \ref{Flo:final} we can see that constrains in $\varepsilon_{\ell\ell}^{eP}$
lead to $V^{2}>5.3\vartheta_{2}^{2}$., while the constrains in $\varepsilon_{\ell\ell}^{uP}$
lead to $V^{2}>21.7\vartheta_{2}^{2}$, and the constrains in $\varepsilon_{\ell\ell}^{dP}$~
($|\varepsilon_{\mu\mu}^{dL}|<0.003)$\textcolor{red}{~} lead to
~ $V^{2}>60\vartheta_{2}^{2}$ . If $\vartheta_{2}$ has its maximum
value of $174.105$ GeV then $V\gtrsim1.3$~TeV. We note also that
by $|\varepsilon_{\mu\mu}^{uL}|<0.003$ we obtain $\left|\vartheta_{2}^{2}-\vartheta_{1}^{2}\right|<0.006V^{2}$
, then for $V\sim1.3$TeV and $\vartheta_{2}=174$ GeV we obtain $142$GeV$<\vartheta_{1}<201$GeV.
We therefore can not predict any hierarchy to the VEV\textquoteright{}s
$\vartheta_{1}$ and $\vartheta_{2}$. Based on that results, we obtain
the following inferior limits for the new Gauge bosons masses:
\begin{eqnarray*}
M_{K_{{I}}} & =&M_{Z'}>610\,\mathrm{GeV},\\
\, M_{K'} & >&613\,\mathrm{GeV},\\
M_{K_{{R}}} & >&740\,\mathrm{GeV}.
\end{eqnarray*}

\section{CONCLUSION}

We presented in this work a procedure to show that models with extended
Gauge symmetries $SU(3)_{C}\times SU(3)_{L}\times U(1)_{X}$ can lead
to neutrino non-standard interactions, respecting the Standard Model
Gauge symmetry $SU(3)_{C}\times SU(2)_{L}\times U(1)_{Y}$, without
spoiling the available experimental data and reproducing the known
phenomenology at low energies. We also have shown that with an assumption
about a mass hierarchy for the Higgs triplets VEV\textquoteright{}s
we could qualitatively address the mass hierarchy problem in standard
model. Finally we obtained limits for the triplets VEV\textquoteright{}s
based on limits for NSI in cross-section experiments. 

We believe that the class of model presented here is an interesting
theoretical possibility to look for new physics beyond SM. We restrained
our work to a simple scenario, but flavor-changing interactions can
be naturally introduced in the model, leading to new constraints on
NSI.
\begin{acknowledgments} 
We would like to thank Alex Dias and Marcelo Guzzo for valuable discussions. 
One of us (M.M.) would like to thank CNPq for financial support.

\end{acknowledgments}

\end{document}